%% file: spatioTemporalDynamicsOfSpeechProduction.tex
\documentclass{article}

\usepackage{hyperref}       % hyperlinks
\usepackage{url}            % simple URL typesetting
\usepackage{amsfonts}       % blackboard math symbols
\usepackage{graphicx}       % microtypography
\usepackage{natbib}
\usepackage[nonumberlist]{glossaries}

\title{Entrainment of traveling waves\\to rhythmic motor acts}

\author{
  Joaqu\'{i}n Rapela\thanks{rapela@ucsd.edu}\\
  Swartz Center for Computation Neuroscience\\
  University of California San Diego
}

\begin{document}

\maketitle

\input{glossary}

\makeglossaries

\begin{abstract}
\input{abstract}
\end{abstract}
\thispagestyle{empty}

\pagebreak

\printglossaries
\thispagestyle{empty}

\pagebreak

\tableofcontents
\thispagestyle{empty}

\pagebreak

\setcounter{page}{1}

\section{Introduction}

\input{introduction}

\section{Results}

\subsection{Phase Alignment}
\label{sec:phaseAlignment}

\input{phaseAlignment}

\subsection{Amplitude Modulations}
\label{sec:amplitudeModulations}

\input{amplitudeModulations}

\subsection{Phase-Amplitude Coupling (PAC)}
\label{sec:pac}

\input{pac}

\subsection{Traveling Waves (TWs)}
\label{sec:tws}

\input{tws}

\subsection{Organization of PAC in the Presence of TWs}
\label{sec:pacWithTWs}

\input{pacWithTWs}

\section{Discussion}

\input{discussion}

\section{Acknowledgments}

\input{acknowledgments}

\pagebreak

\small

\bibliographystyle{plainnatNoNote}
\bibliography{rhythms,speech,eeg,stats,pac,travelingWaves,surfaceLaminar,vision,brainImaging}

\setcounter{page}{1}

\renewcommand\thefigure{\thesection.\arabic{figure}}    
\setcounter{figure}{0}

\renewcommand\thetable{\thesection.\arabic{table}}    
\setcounter{table}{0}

\clearpage
\appendix
% \numberwithin{equation}{section}

\input{supplementaryInfo}

\end{document}

%% file: glossary.tex
\newglossaryentry{PAC} {
    name={PAC},
    description={phase-amplitude coupling. Coupling between the phase of
low-frequency oscillations and the amplitude of high-frequency ones
(Sections~\ref{sec:pac} and~\ref{sec:pacCurveAndMI})}
}

\newglossaryentry{ERP} {
    name={ERP},
    description={event-related potential. Mean of voltages recorded at a single
electrode and aligned with respect to an event of interest (e.g., the
transition time between a consonant and a vowel).}
}

\newglossaryentry{MI} {
    name={MI},
    description={modulation index quantifying the strength of phase-amplitude coupling (Sections~\ref{sec:pac} and~\ref{sec:pacCurveAndMI})}
}

\newglossaryentry{TW} {
    name={TW},
    description={traveling wave (Section~\ref{sec:tws})}
}

\newglossaryentry{CV} {
    name={CV},
    description={consonant vowel}
}

\newglossaryentry{vSMC} {
    name={vSMC},
    description={ventral sensorimotor cortex. A brain region that controls the
vocal articulators}
}

\newglossaryentry{ECoG} {
    name={ECoG},
    description={electrocorticography. A brain recording modality that measures
potentials directly from the cortical surface}
}

\newglossaryentry{MEG} {
    name={MEG},
    description={magnetoencephalography. A brain recording modality that 
measures magnetic fields from the scalp.}
}

\newglossaryentry{EEG} {
    name={EEG},
    description={electroencephalography. A brain recording modality that 
measures electrical potentials from the scalp.}
}

\newglossaryentry{ITC} {
    name={ITC},
    description={inter-trial coherence. A measure of phase alignment between
trials (Section~\ref{sec:itc})}
}

\newglossaryentry{ERSP} {
    name={ERSP},
    description={event-related spectral perturbation. An average measure of evoked power across trials (Section~\ref{sec:ersp})}
}

%% file: abstract.tex
A remarkable early observation on brain dynamics is that when humans are
exposed to rhythmic stimulation their brain oscillations entrain to the rhythm
of the stimuli~\citep{adrianAndMatthews34}. Currently it is not know whether
rhythmic motor acts can entrain neural oscillations.
We investigated this possibility in an experiment where a subject produced
consonant-vowel (\gls{CV}) syllables in a rhythmic fashion, while we performed
ECoG recordings with a dense grid covering most speech processing brain regions
across the left hemisphere.
Most strongly over the ventral sensorimotor cortex (\gls{vSMC}), a cortical
region that controls the vocal articulators, we detected significant
concentration of phase across trials at the specific frequency of speech
production. We also observed amplitude modulations. In addition we found
significant coupling between the phase of brain oscillations at the frequency
of speech production and their amplitude in the high-gamma range (i.e.,
phase-amplitude coupling, \gls{PAC}).  Furthermore, we saw that brain
oscillations at the frequency of speech production organized as traveling waves
(\glspl{TW}), synchronized to the rhythm of speech production.
It has been hypothesized that \gls{PAC} is a mechanism to allow low-frequency
oscillations to synchronize with high-frequency neural activity so that spiking
occurs at behaviorally relevant times. If this hypothesis is true, when
\gls{PAC} coexists with \glspl{TW}, we expect a specific organization of
\gls{PAC} curves.  We observed this organization experimentally and verified
that the peaks of high-gamma oscillations, and therefore spiking, occur at the
same times across electrodes. Importantly, we observed that these spiking times
were synchronized with the rhythm of speech production.
To our knowledge, this is the first report of motor actions organizing (a) the
phase coherence of low-frequency brain oscillations, (b) the coupling between
the phase of these oscillations and the amplitude of high-frequency
oscillations, and (c) \glspl{TW}. It is also the first demonstration that
\glspl{TW} induce an organization of \gls{PAC} so that spiking across spatial
locations is synchronized to behaviorally relevant times.

%% file: introduction.tex
External visual~\citep{regan66} and auditory~\citep{galambosEtAl81} rhythmic
stimuli entrain brain oscillations (i.e., drag brain oscillations to follow the
rhythm of the stimuli), and this entrainment is modulated by attention so that
the occurrence of attended stimuli coincides with the phase of brain
oscillations of maximal
excitability~\citep{lakatosEtAl05,lakatosEtAl08,lakatosEtAl13,oconnellEtAl11,besleEtAl11,gomezRamirezEtAl11,zionGolumbicEtAl13,cravoEtAl13,mathewsonEtAl10,spaakEtAl14,grayEtAl15}.
% references from zoefelAndVanRullen15
Although speech is only quasi rhythmic~\citep{cummins12} an increasing number of
studies is showing that neural oscillations can entrain to speech
% sound~\citep{dingAndSimon12a,dingAndSimon12b,dingAndSimon13,dingAndSimon14,peelleAndDavis12,zionGolumbicEtAl12,zionGolumbicEtAl13b,dingEtAl13,grossEtAl13,hortonEtAl13,peelleEtAl13,powerEtAl13,steinschneiderEtAl13,doellingEtAl14,millmanEtAl15,parkEtAl15}.
sound~\citep[e.g.,][]{zionGolumbicEtAl13b,grossEtAl13,doellingEtAl14,millmanEtAl15,parkEtAl15}.
For recent reviews on the entrainment of neural oscillations to speech sound
see \citet{peelleAndDavis12,zionGolumbicEtAl12,dingAndSimon14}.  What is
currently not know if whether speech production can entrain neural
oscillations.  Here we present evidence showing that a speaker rhythmically
producing consonant-vowel (\gls{CV}) syllables at different frequencies
entrains brain oscillations in the ventral sensorimotor cortex~(\gls{vSMC}) at
corresponding different frequencies.

It has been suggested that a role of entrained low-frequency oscillations is to
modulate the excitability of neurons, in such a way that periods of higher
excitability correspond to events of interest in sensory
streams~\citep{schroederAndLakatos09}. This hypothesis has been supported by
reports of phase-amplitude coupling (\gls{PAC}), where the phase of
low-frequency oscillations modulates the amplitude of higher-frequency
ones~\citep[e.g.,][]{canoltyEtAl06}. In speech perception it has been proposed
that theta rhythm (3-8~Hz) is related to the encoding of slower syllabic
information in the speech signal, the gamma rhythm (\textgreater 30~Hz) is involved in the
linguistic coding of phonemic details, and the \gls{PAC} between the theta and
gamma rhythms could modulate the excitability of neurons to devote more
processing power to the informative parts of syllabic sound
patterns~\citep{giraudAndPoeppel12,hyafilEtAl15}.  However, \gls{PAC} has not
yet been reported in speech-production brain regions. We show that over the
\gls{vSMC} the low-frequency oscillations entrained to the rhythm of speech
production are strongly coupled with high-gamma oscillations. Importantly, we
demonstrate that the phase at which the entrained oscillation couples with the
maximum amplitude of the high-gamma oscillations changes orderly from ventral
to dorsal electrodes over the \gls{vSMC}.

Traveling waves (\glspl{TW}) have been reported in animal
studies~\citep[e.g.,][]{rubinoEtAl06}, human
\gls{EEG}~\citep[e.g.,][]{pattenEtAl12}, and human
\gls{ECoG}~\citep[e.g.,][]{bahramisharifEtAl13}. However, these waves have not
been observed in speech production brain regions in humans. The aforementioned
orderly change of \gls{PAC} across the \gls{vSMC} suggested the existence of
\glspl{TW}. We looked for them and we found them.  As mentioned above, brain
activity at single recording sites can be entrained to external stimulation
but, to our knowledge, it is unknown whether distributed spatio-temporal
activity in the form of \glspl{TW} can also be entrained. We demonstrate that
\glspl{TW} over the \gls{vSMC} are entrained to the rhythm of speech
production.

\gls{PAC} has been studied in the absence of \glspl{TW}. When these waves are
present, in order for cells to fire at the same behaviorally-relevant time,
\gls{PAC} curves measured across electrodes should be different from each other
and systematically organized.  If these curves were identical to each other,
cells would fire in the order given by the propagating \gls{TW}, and not at the
same time.  Below we propose an organization of \gls{PAC} so that when
\glspl{TW} are present cells fire at the same time, we show that this
organization is present in our recordings, and provide evidence indicating that
cells along the direction of propagation of a planar \gls{TW} fire at the same
behaviorally relevant time.

%% file: phaseAlignment.tex
\begin{figure}
\begin{center}
\includegraphics[width=5in]{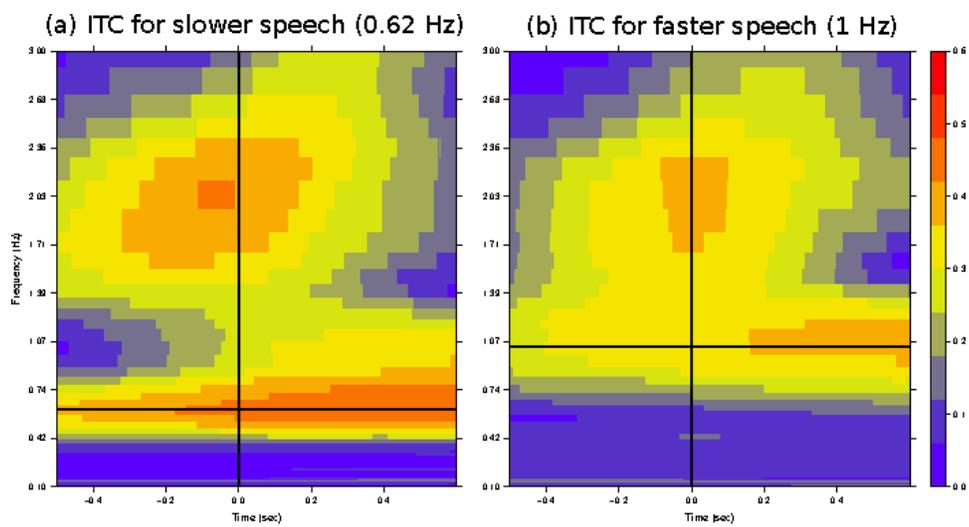}
\end{center}

\caption{\glspl{ITC} for electrode 154 computed from recordings in experimental
session EC2\_B105 with an inter-syllable production frequency of 0.62~Hz (a),
and from experimental session EC2\_B89 with an inter-syllable production
frequency of 1.03~Hz (b). Note that the peaks of the \gls{ITC} at the lowest
frequencies (red points) align with the median frequency of speech production
(black horizontal line) both when the subject speaks slower (a) and faster
(b).}

\label{fig:itc154}
\end{figure}
\begin{figure}
\begin{center}
\includegraphics[width=5.0in]{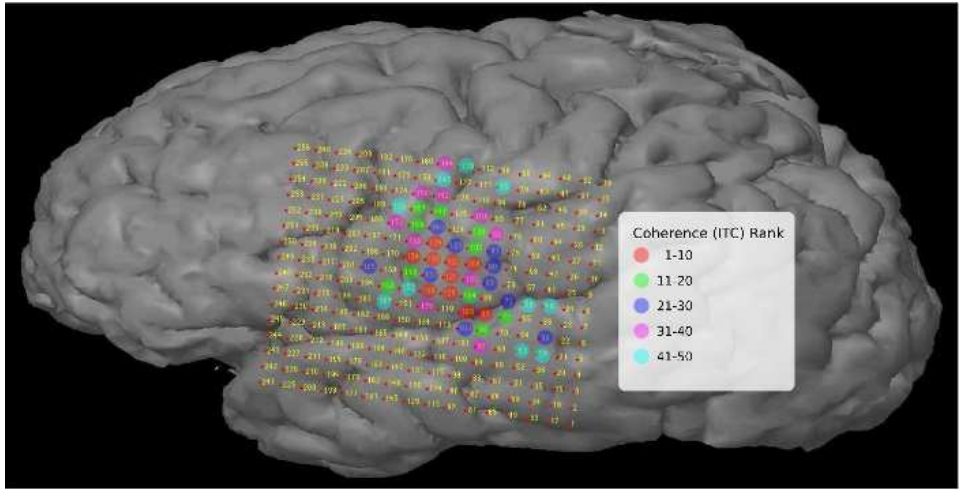}
\end{center}

\caption{Strongest entrainment occurs over the \gls{vSMC}. The 50 electrodes with
largest \gls{ITC} (between 0.5 seconds before and 0.6 seconds after the \gls{CV}
transition, at the frequency of \gls{CV} production, in the experimental session
EC2\_B105) are highlighted in color.}

\label{fig:maxITCAcrossElectrodesB105}
\end{figure}

We epoched the raw \gls{ECoG} waveforms around the time of \gls{CV} transitions
(represented by the vertical line at time zero in all figures). We analyzed
separately recordings from different sessions, and present the analysis from
recording sessions EC2\_B105 and EC2\_B89, the sessions with the largest (1.62
sec) and shortest (0.97 sec) median inter-syllable separation times,
respectively (Figure~\ref{fig:issHist}).
Figure~\ref{fig:itc154}a and~\ref{fig:itc154}b show the inter-trial coherence
(\gls{ITC}, Section~\ref{sec:itc}), a measure of phase alignment in a group of
trials, calculated from recordings in sessions EC2\_B105 and EC2\_B89,
respectively, for electrode 154 in the \gls{vSMC}
(Figure~\ref{fig:maxITCAcrossElectrodesB105}).  In agreement with our
hypothesis that the rhythm of speech production entrains low-frequency brain
oscillations, in the \gls{ITC} computed from data from experimental session
EC2\_B105 with a median inter-syllable production frequency of 0.62~Hz (black
horizontal line in Figure~\ref{fig:itc154}a), we observe a peak of \gls{ITC} around
this frequency (red pixels in Figure~\ref{fig:itc154}a).  Also, we see a peak
of \gls{ITC} around a frequency of 1.03~Hz in Figure~\ref{fig:itc154}b, obtained
from an experimental session where the median inter-syllable separation time
was 1.03~Hz.  These figures illustrate that low-frequency brain oscillations
entrain to the rhythm of speech production.

For each electrode, we computed the maximum \gls{ITC} between 0.5 seconds before and
0.6 seconds after the \gls{CV} transition at the frequency of speech production.
Figure~\ref{fig:maxITCAcrossElectrodesB105} highlights the 50
electrodes with largest \glspl{ITC}. We sorted the electrodes by their maximum \gls{ITC}
value, and colored them according to their rank in this sorting. The ten
electrodes with largest \gls{ITC} are colored in red and are located in the center
of the ventral sensoritmotor cortex. As the distance of an electrode to the
center of the \gls{vSMC} increases, its maximum \gls{ITC} decreases.
Section~\ref{sec:itcsAcrossvSMC} more clearly
illustrates the specificity of phase coherence to the \gls{vSMC} by showing
\gls{ITC}
plots from the ventral to the dorsal edge of the grid across the \gls{vSMC}.

%% file: amplitudeModulations.tex
\begin{figure}
\begin{center}
\includegraphics[width=5in]{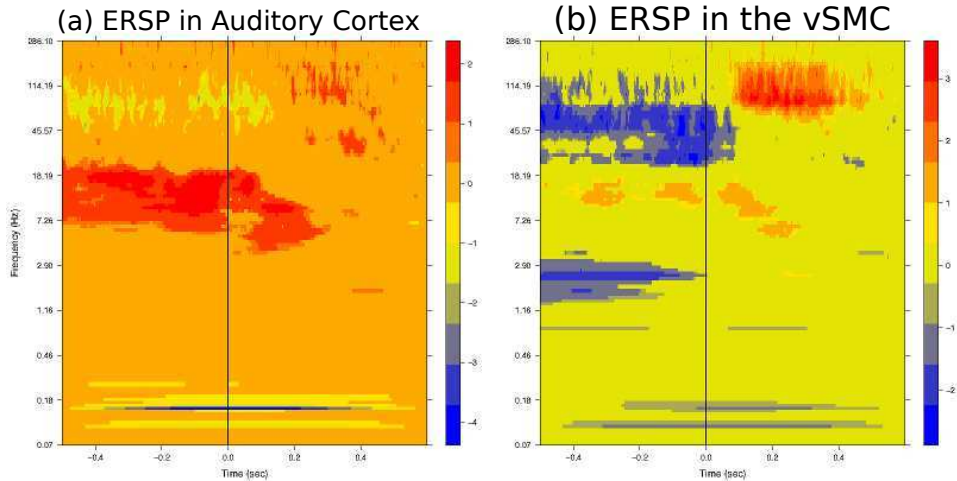}
\end{center}

\caption{\glspl{ERSP} for electrode 133 in the auditory cortex (a) and for electrode
135 in the \gls{vSMC} (b). In the auditory cortex (a) we observe a peak of amplitude
(red blob) in the beta range before the \gls{CV} transition (vertical black line),
while in the \gls{vSMC} (b) a peak of amplitude occurs in the high-gamma range after
the \gls{CV} transition.}

\label{fig:erspsAuditoryAndVSMC}
\end{figure}

The Event-Related Spectral Perturbation (\gls{ERSP}, Section~\ref{sec:ersp}) is
a measure of power modulations around an event of interest.
Figure~\ref{fig:erspsAuditoryAndVSMC}a shows \glspl{ERSP} around the transition
time between consonants and vowels (black vertical line in
Figure~\ref{fig:erspsAuditoryAndVSMC}).
In electrode 133 over auditory cortex (Figure~\ref{fig:erspsAuditoryAndVSMC}a)
we see a significant power increase (red blob) before the transition between
consonants and vowels in the beta range.  Differently, in electrode 135 over
the \gls{vSMC} (Figure~\ref{fig:erspsAuditoryAndVSMC}b) we observe a
significant power increase after the transition between consonants and vowels
and in the high-gamma range.

Figures~\ref{fig:ersp129B105}-\ref{fig:ersp141B105} show \glspl{ERSP} between the ventral electrode 129 and the dorsal
electrode 141 in experimental
session EC2\_B105.
Power in the beta range increases from electrodes 129 to 132
(Figures~\ref{fig:ersp129B105}-\ref{fig:ersp132B105}) in the ventral temporal
cortex, and peaks over electrodes 133 and 134 around the auditory cortex
(Figures~\ref{fig:ersp133B105} and~\ref{fig:ersp134B105}), prior to
consonant-vowel transitions. A sharp transition is seen in
electrode 135 over the ventral sensorimotor cortex
(Figure~\ref{fig:ersp135B105}), with strong modulations in the high-gamma range
(70-200~Hz), after consonant-vowel transitions, and almost no modulation in
the beta range. This pattern of modulations is strongest over electrode 136
(Figure~\ref{fig:ersp136B105}), and fades along electrodes 137-141
(Figures~\ref{fig:ersp137B105}-\ref{fig:ersp141B105}) toward the dorsal end of
the \gls{vSMC}.

%% file: pac.tex
\begin{figure}
\begin{center}
\includegraphics[width=2.5in]{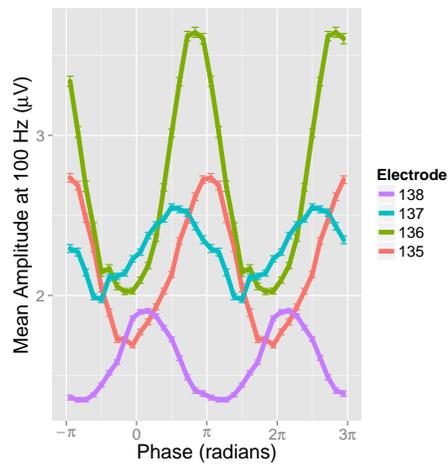}
\end{center}

\caption{Phase-amplitude coupling curves at electrodes over the \gls{vSMC}. Phase and
amplitude were measured at the entrainment frequency (0.62~Hz) and at 100~Hz,
respectively. All displayed electrodes showed a strong modulation of the
high-gamma amplitude by the entrained low-frequency phase.  Note that as we
move from more ventral to more dorsal electrodes (i.e., from electrode 135 to
electrode 138) the peak of the \gls{PAC} curves occurs at earlier phases. The
significance of this organization is discussed in
Section~\ref{sec:pacWithTWs}.}

\label{fig:pac}
\end{figure}

For experimental session EC2\_B105, we studied the coupling between phase at
the entrainment frequency of 0.62~Hz (0.01~Hz, wavelet full width at half
maximum) and amplitude at 100~Hz (2.06~Hz, wavelet full width at half
maximum).  For each electrode we calculated the \gls{PAC} curve
(Section~\ref{sec:pacCurveAndMI}), and quantified the strength of the coupling
using the modulation index (\gls{MI}, Section~\ref{sec:pacCurveAndMI}).
Figure~\ref{fig:pac} shows \gls{PAC} curves measured at electrodes 135-138 (for
visualization purposes we show two phase cycles). The \gls{ECoG} power at 100~Hz is
well modulated by the phase at the entrainment frequency, with \glspl{MI} 4.84e-3,
8.22e-3, 1.05e-3, and 2.82e-3 at electrodes 135-138, respectively.  For
electrode 136 (green trace) larger \gls{ECoG} amplitudes in the high-gamma range
(100~Hz) tend to occur at the valley (phase near $\pi$) of the low-frequency
oscillation (0.62~Hz).  As we move from the more ventral electrode 135 to the
more dorsal electrode 138, the phase of the low-frequency oscillation at which
the amplitude of the high-gamma oscillations is maximal shifts gradually from
$\pi$ (pink trace) to 0 (violet trace). We discuss this organization in
Section~\ref{sec:pacWithTWs}.
Strongest \gls{PAC} between phase at the entrainment frequency and amplitude at
100~Hz was observed over the \gls{vSMC} and surrounding regions.
Figure~\ref{fig:largestMIsB105} highlights the 50 electrodes with largest
\gls{MI}.  

\begin{figure}
\begin{center}
\includegraphics[width=5.5in]{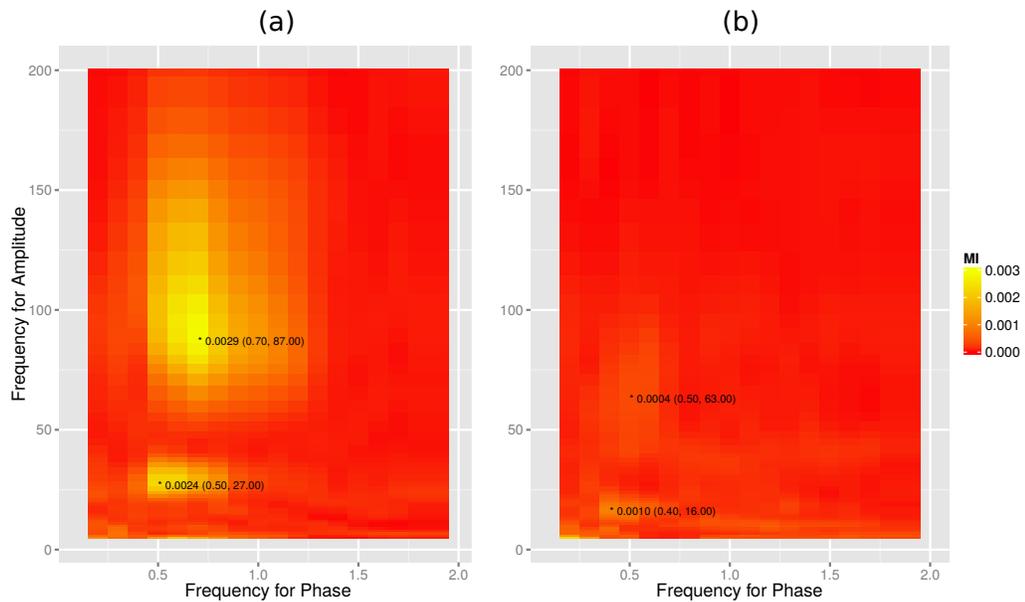}
\end{center}

\caption{Modulation indices for electrode 154 over the \gls{vSMC} (a) and for
electrode 132 over the auditory cortex (b) at multiple frequencies for phase
and amplitude.
Over the \gls{vSMC} (a) the peak \gls{MI} for amplitudes in the high-gamma
range occurs at the entrained frequency for phase (0.62~Hz), while the peak
\gls{MI} for amplitudes in the beta range occurs at lower frequencies. This
suggests that over the \gls{vSMC} not only phase coherence, but also
phase-amplitude coupling, is matched to the speech production frequency.
Over auditory cortex (b) \glspl{MI} are much smaller than over the \gls{vSMC}
and the peak \glspl{MI} occurred at frequencies for phase below the entrainment
frequency. This suggests that \gls{PAC} is largest over speech-production
brain regions and that the maximization of \gls{PAC} at the entrainment
frequency is specific to the \gls{vSMC}.}

\label{fig:misForMultipleFreqsB105}
\end{figure}

For electrode 154 in the \gls{vSMC}, we computed the \gls{MI} for all pairs of
frequency for phase and frequency for amplitude, with the frequency for phase
between 0.1 and 20~Hz, and the frequency for amplitude between 5 and 200~Hz. We
only observed large \glspl{MI} for frequencies for phase between 0.1 and 2~Hz.
Figure~\ref{fig:misForMultipleFreqsB105}a plots the obtained \glspl{MI}.  We
found two clusters of the frequency for amplitude with large \glspl{MI}. The
first cluster was on the high-gamma range (60 to 150~Hz) and the second cluster
on the beta range (20 to 40~Hz).  An asterisk in
Figure~\ref{fig:misForMultipleFreqsB105} marks a \gls{MI} peak, the number next
to the asterisk is the peak \gls{MI}, and the first and second numbers in the
parenthesis are the frequency for phase and for amplitude, respectively, of the
peak \gls{MI}.  Interestingly, for the cluster on the high-gamma range, the
maximum \gls{MI} occurred exactly at the bin of the entrainment frequency
(0.62~Hz for the frequency for phase).  Differently, for the cluster in beta
range the maximum \gls{MI} occurred below the entrainment frequency (0.5~Hz for
the frequency for phase). This figure suggests that not only low-frequency
oscillations become entrained to the rhythm of speech production, but also
\gls{PAC} is optimized at this rhythm.  For electrode 132 in the auditory
cortex, Figure~\ref{fig:misForMultipleFreqsB105}b, the \glspl{MI} were overall
much weaker, and both the clusters in the high-gamma and beta ranges showed a
peak \gls{MI} at frequencies for phase below the entrainment frequency. This
suggest that the maximization of \gls{PAC} at the entrainment frequency is
specific to the \gls{vSMC}.

%% file: tws.tex
\begin{figure}
\begin{center}
\includegraphics[width=5in]{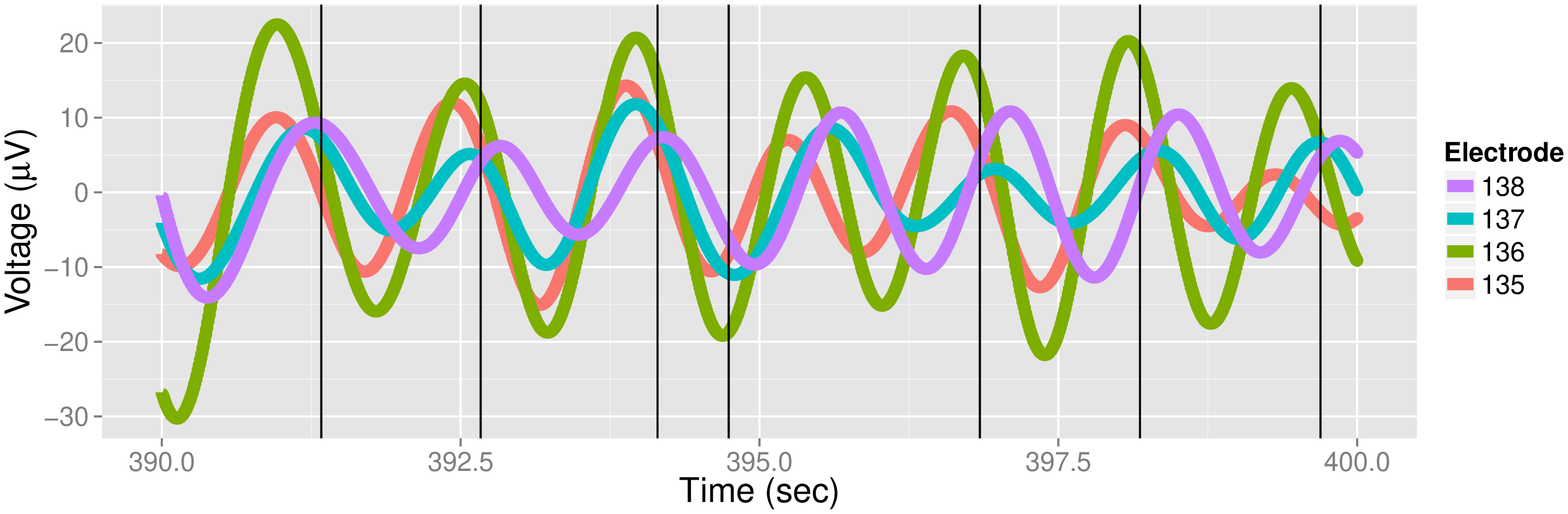}
\end{center}

\caption{Filtered voltage waveforms along the ventral-dorsal axis of the
\gls{vSMC}.
Filtering was performed between 0.4 and 0.8~Hz, around the median frequency of
\gls{CV} syllable production of 0.62~Hz. Vertical black lines indicate
\gls{CV} transition
times. As we move from the ventral electrode 135 in red to the dorsal electrode
141 in magenta, voltage waveforms are orderly shifted to right, indicating the
existence of a traveling wave. Also, the peak of the waves occurs around the
time of \gls{CV} transitions, suggesting that these traveling waves are entrained to
the rhythm of speech production.}

\label{fig:tws}
\end{figure}

We filtered the \gls{ECoG} recordings in experimental session EC2\_B105
(Figure~\ref{fig:issHist}) with a second-order bandpass Butterworth filter
around the median syllable-production syllable of 0.62~Hz (low- and
high-frequency cutoffs 0.4~Hz and 0.8~Hz, respectively).  Figure~\ref{fig:tws}
shows the filtered voltages from electrodes 135-141 at the center of the
\gls{vSMC} in a representative time interval (390 to 400~seconds after the
start of the experiment). We see that, as we move from the ventral electrode
135 to the dorsal electrode 141, the voltage traces are orderly shifted to
later times.  The vertical black lines indicate the transition time between a
consonant and a vowel in the production of consonant-vowel syllables. Note that
the peaks of the voltage traces tend to occur around these consonant-vowel
transition times.  Thus, this figure illustrates a wave of activity traveling
from the ventral to the dorsal edge of the ventral sensorimotor cortex and
entrained to the rhythm of speech production.

A movie showing the cosine of the phase of the bandpassed filtered voltages
across the whole grid simultaneously with the speech of the subject can be
found at \url{https://youtu.be/dXrzj2eEuVY}. The bottom/top/left/right pixels
in the movie correspond to electrodes along the ventral/dorsal/frontal/caudal
sides of the grid.
We see mostly planar traveling waves moving in different directions, but
occasionally we also observe rotating traveling waves. At most times when the
subject produces a \gls{CV} syllable we see a white blob over the \gls{vSMC}, indicating
that filtered voltages approach their peak values. This movie demonstrates,
for the first time, the existence of traveling waves across left speech
processing brain areas synchronized to the rhythm of speech production.

%% file: pacWithTWs.tex
If we wanted to design a brain so that all neurons recorded by the four
electrodes in Figure~\ref{fig:pac} fired at the same behaviorally relevant time
(e.g., the transition between the consonant-vowel syllable at time 391.2
seconds), how would be choose the \gls{PAC} curves for these electrodes? Since
the phase at electrode 135 (red trace in Figure~\ref{fig:pac}) at time 391.2 is
close to $\pi$/4 we would use a \gls{PAC} curve for this electrode with maximal
high-gamma amplitude at this phase. Then, because high-gamma amplitude is
related to neural firing~\citep{rayEtAl08}, cells around electrode 135 would
then tend to spike at the time of the consonant vowel transition.  Next, due to
the traveling wave, the phase at electrode 136 at time 391.2 is earlier than
that at electrode 135, thus we would use a \gls{PAC} curve with a peak at this
earlier phase for electrode 136.  Similarly, we would use a \gls{PAC} curve
with an earlier peak at electrode 137 than that at electrode 136. That is, as
we move along the direction of propagation of a traveling wave, we would use
\gls{PAC} curves with earlier and earlier peaks.  And this is what we observed
experimentally in Figure~\ref{fig:pac}.

\begin{figure}
\begin{center}
\includegraphics[width=5.0in]{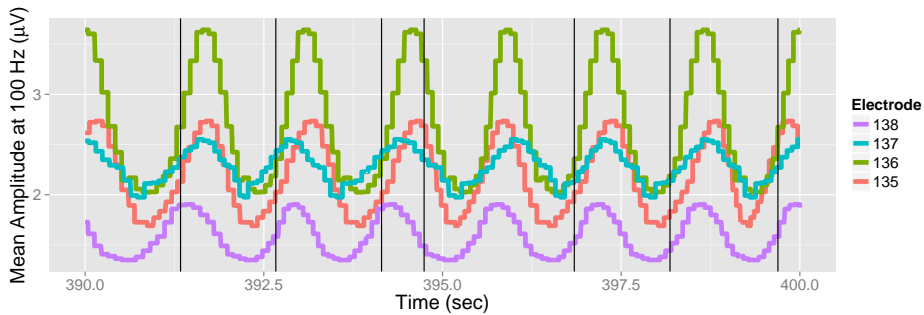}
\end{center}

\caption{Mean high-gamma amplitude as a function of time at electrodes over the
ventral sensorimotor cortex. Each time point at each electrode corresponds to a
given phase of the oscillation at the speech-production frequency of 0.62~Hz in
Figure~\ref{fig:tws}, and from Figure~\ref{fig:pac} each phase corresponds
to a given mean amplitude at 100~Hz. Thus, each time point of each electrode
corresponds to a mean phase at 100~Hz. Colored traces show this correspondence
across different electrodes. We observe that mean high-gamma amplitude, and
therefore spiking, is aligned across electrodes over the ventral sensorimotor
cortex and synchronized to the rhythm of speech production. The curves are not
smooth because we used coarse bins in the calculation of the \gls{PAC} curves in
Figure~\ref{fig:pac}.}

\label{fig:amplitudeVsTime}
\end{figure}

To validate that the previous organization of \gls{PAC} curves allows cells to
fire at the same behaviorally relevant time in the presence of traveling waves,
Figure~\ref{fig:amplitudeVsTime} plots mean high-gamma amplitude as a function
of time for electrodes 135-138 (for each time and each electrode, we extracted
the phase at 0.62~Hz, then the mean amplitude at 100~Hz for this phase from the
\gls{PAC} curve in Figure~\ref{fig:pac}, and finally we plotted the extracted
mean amplitudes as a function of the corresponding times). As expected, we
found that high-gamma amplitude, and therefore cell spiking, peaks at
approximately similar times across electrodes. In addition, we observed that
the high-gamma amplitude peaks occur shortly after the transition time between
consonant and vowels, which shows that the traveling waves are therefore
entrained to the rhythm of speech production.

%% file: discussion.tex
We have shown that rhythmic speech production entrains low-frequency brain
oscillations (Figure~\ref{fig:itc154}) and that this entrainment is stronger
over the \gls{vSMC} (Figure~\ref{fig:maxITCAcrossElectrodesB105}), that
rhythmic speech production modulates the power of beta oscillations over the
auditory cortex and that of high-gamma oscillations over the \gls{vSMC}
(Figure~\ref{fig:erspsAuditoryAndVSMC}), that the phase of the entrained
low-frequency oscillations couples with the amplitude of high-gamma
oscillations (Figure~\ref{fig:pac}) and that this coupling is largest over the
\gls{vSMC} (Figure~\ref{fig:largestMIsB105}), that over the \gls{vSMC} --but
not over the auditory cortex-- this coupling is maximal between phase at the
entrained frequency and amplitude in the high-gamma range
(Figure~\ref{fig:misForMultipleFreqsB105}), and that the entrained oscillations
organize as \glspl{TW} synchronized to the produced speech
(Figure~\ref{fig:tws} and movie at \url{https://youtu.be/dXrzj2eEuVY}).
Finally, we observed that the peaks of \gls{PAC} curves occur at earlier phases
along the direction of propagation of the \glspl{TW} and presented evidence
(Figure~\ref{fig:amplitudeVsTime}) indicating that this organization of
\gls{PAC} curves enables neurons along the direction of propagation of the
\gls{TW} to fire at the same behaviorally relevant time.

These initial findings open several interesting directions for future
investigations. To mention a few:
(1) How do entrainment, \gls{PAC}, and \glspl{TW} relate to properties of the
produced \gls{CV} syllables? For example, does \gls{PAC} occur earlier for
fricatives than for nasals?
(2) How do these effects (i.e., entrainment, \gls{PAC}, and \glspl{TW})
correspond to the specific articulators recruited in the production of a
\gls{CV} syllable? For example, do planar \glspl{TW} correspond to the
recruitment of the jaw but rotating \glspl{TW} to the recruitment of the
larynx?
(3) \gls{PAC} has been reported in speech
perception~\citep[e.g.,][]{giraudAndPoeppel12,hyafilEtAl15}, but not in speech
production tasks. We claimed that the \gls{PAC} we observed is related to
speech production, but we have not eliminated the possibility that it may be
related to the perception of the speaker's own speech. The detection of
\gls{PAC} in new experiments eliminating speech feedback would confirm that
\gls{PAC} can be related to the production of speech.
(4) The observed effects appeared in the strongly rhythmic production of
\gls{CV} syllables. Would similar effects be observed in colloquial speech? How
do prosodic features of speech, like intonation or stress, affect the observed
effects?
(5) We have characterized phase coherence and \gls{PAC} at single electrodes,
but probably phase coherence and \gls{PAC} across different electrodes play an
important role in the synchronization and information transfer across different
speech perception and production brain regions, as observed in
non-speech-processing tasks~\citep[e.g.,][]{voytekEtAl15}.

It is possible to explain some of our results in terms of event-related
potentials (\glspl{ERP}, the mean recorded potential around an event of
interest).  Previous studies have suggested that \glspl{ERP} are generated by
the alignment of phases of ongoing oscillations~\citep{makeigEtAl02}, but this
suggestion is still controversial~\citep{sausengEtAl07}. Because the production
of \gls{CV} syllables occurred at a low-frequency rhythm, \glspl{ERP} should be
generated at this low-frequency and we should observe alignment of phases
at this low frequency.  Thus, this \gls{ERP} argument explains
Figure~\ref{fig:itc154}.
Also, since \glspl{ERP} should be stronger over the \gls{vSMC}, we should
expect more phase alignment over this region, explaining
Figure~\ref{fig:maxITCAcrossElectrodesB105}. One could argue that the
production of speech is related to spiking, and therefore to high-gamma
activity, over the \gls{vSMC}. Thus, the low frequency phase at which
oscillations reset for the generation of \glspl{ERP} should be related to
high-gamma activity. This argument addresses Figures~\ref{fig:pac}
and~\ref{fig:misForMultipleFreqsB105}. 

Section~\ref{sec:erpsAcrossvSMC} shows \glspl{ERP} computed along the
ventral-dorsal axis of the recording grid. In agreement with
Figure~\ref{fig:maxITCAcrossElectrodesB105}, we see strongest \glspl{ERP} at
electrodes 136-140 over the \gls{vSMC}. These \glspl{ERP} begin around 150~msec
before the \gls{CV} transition, peak around 100~ms after this transition, and
show a negative peak around 400~ms after the \gls{CV} transition.  Also, in agreement
with Figure~\ref{fig:pac} we see that the peak of the \glspl{ERP} are shifted
to later times as we move from electrode 136 to electrode 140, also suggesting
the existence of \glspl{TW}.

Thus, the spectral and \gls{ERP} analysis are consistent with each other. An
advantage of the former is that it decomposes the averaged effects seen in
\glspl{ERP} in different frequency bands. Since several frequency-specific
effects on electrophysiological recordings are known, spectral analysis allow
richer descriptions of electrophysiological recordings than \gls{ERP} ones. For
instance by decomposing averaged effects in different frequency bands, our
spectral analysis showed that neural spiking, as reflected by high-gamma
amplitude, was coupled to relevant times in the speech production rhythm, as
indicated by the coupled phase at the speech-production frequency
(Figure~\ref{fig:pac}).
Also, our \gls{PAC} analysis is more quantitative than the previous argument
linking \glspl{ERP} and high-gamma amplitude. This analysis revealed a new
organization of \gls{PAC} in the presence of \glspl{TW}, providing strong
support to a previous hypothesis stating that, when stimuli or behavior occurs
in a rhythmic fashion, \gls{PAC} is a mechanism that allows neurons to spike at
behavioral relevant times \citep{canoltyAndKnight10}.
In summary, the spectral methods used above to characterize rhythmic speech
production are consistent with, but superior than, \gls{ERP} methods for the
characterization of rhythmic behaviors such as speech.

The effects reported above reflect brain mechanisms beyond the control of vocal
articulators. For example, the traveling wave movie
(\url{https://youtu.be/dXrzj2eEuVY}) shows waves moving across
multiple speech processing brain regions. At several times we see waves moving
from electrodes over the auditory cortex to electrodes over the \gls{vSMC},
suggesting that these waves could be a mechanism for integrating information
across the brain, as suggested by
others~\citep[e.g.,][]{nunezAndSrinivasan06b,satoEtAl12}.

As speaking must be compatible with the performance limits of motor systems,
several theories have linked functional properties of motor systems to speech
capability. The Frame/Content theory~\citep{macNeilageAndDavis00,macNeilageAndDavis01} proposes that the features of elementary production
units, i.e., syllables, are determined by mechanical properties of the speech
apparatus, e.g., natural oscillatory rhythms~\citep{giraudEtAl07}. Previous
investigations~\citep[e.g.,][]{morillonEtAl10} have hypothesized that motor
areas express low-frequency oscillatory activity characteristic of jaw
movements (4~Hz) and high-frequency activity corresponding to movements of the
tongue (e.g., trill at 35-40~Hz). However, to our knowledge this is the first
study providing physiological evidence for the relation of low- and
high-frequency oscillations, as well as their interactions, to speech
production.

Natural speech production is substantially more complex than the production of
\gls{CV} syllables, which suggests that the findings reported in this article
may not be relevant to natural speech. We argue on the contrary. \gls{CV}
syllables are the building blocks of natural speech, such as sinusoidal grating
are the building blocks of natural images. While producing \gls{CV} syllables
subjects spontaneously enter a strongly rhythmic behavioral state.  Natural
speech is strongly rhythmic and the production of \gls{CV} syllables could be
ideal to isolate rhythmic aspects of this behavior.  By characterizing
responses of visual cells to sinusoidal gratings, it has been possible to
understand key neural mechanisms of visual processing, like contrast grain
control~\citep{ohzawaEtAl85}, that are difficult to observe in responses of
visual neurons to complex natural images.  Similarly, investigating the
production of \gls{CV} syllables may be instrumental in discovering key
rhythmic mechanisms of speech production that are obscured in more complex
speech-production tasks.

The specific attributes of mesoscopic cortical activity related to speech
production are not well characterized. In speech perception,
\citet{luoAndPoeppel07} have shown that phase activity (over the human auditory
cortex in the theta range (4-8~Hz) recorded with \gls{MEG}) reliably tracks and
discriminates spoken sentences and that this discrimination is correlated with
speech intelligibility. Here we extended these results and showed that phase
activity (in the form of phase alignment, \gls{PAC}, and \glspl{TW}) is also a
relevant mesoscopic attribute to characterize the neurophysiology of speech
production.

Having detected phase alignment, \gls{PAC} and \glspl{TW} in \gls{ECoG}
recordings, one could use forward models to calculate how these effects
propagate to scalp-recorded potentials in the \gls{EEG} or magnetic fields in
the \gls{MEG}. Next, one could invert these forward models to infer the
existence and properties of these oscillatory effects from \gls{EEG} or
\gls{MEG} recordings~\citep{fristonEtAl07}. If successful this approach would
allow to detect phase alignment, \gls{PAC}, and \glspl{TW} from non-invasive
recordings, and provide novel measures to non-invasively characterize speech
production in health and disease.

%% file: acknowledgments.tex
We thank Dr.~Edward Chang and Dr.~Kristofer Bouchard for sharing the \gls{ECoG}
recordings with us and for comments on a preliminary version of this
manuscript, and Dr.~John Iversen for suggesting us the relation between
\gls{PAC} and \glspl{TW}.

%% file: supplementaryInfo.tex
\section{Supplementary Information}

\subsection{Methods}

\subsubsection{Circular statistics concepts}
\label{sec:circularStats}

This section introduces concepts from circular
statistics \citep{mardia72} used below
to define \gls{ITC}.
Given a set of circular variables (e.g., phases), $\theta_1, \ldots,
\theta_N$, we associate to each circular variable a two-dimensional unit
vector. Using notation from complex numbers, the unit vector associated with
variable $\theta_i$ is:

\begin{eqnarray}
vec(\theta_i)=e^{j\theta_i}
\label{eq:unitVector}
\end{eqnarray}

\noindent The \emph{resultant vector}, $\mathbf{R}$, is the sum of the associated unit
vectors: 

\begin{eqnarray}
\mathbf{R}(\theta_1, \ldots, \theta_N)=\sum_{i=1}^{N}vec(\theta_i)
\label{eq:resultantVector}
\end{eqnarray}

\noindent The \emph{mean resultant length}, $\bar{R}$, is the length of the resultant
vector divided by the number of variables:

\begin{eqnarray}
\bar{R}(\theta_1, \ldots, \theta_N)=\frac{1}{N}|\mathbf{R}(\theta_1, \ldots,
\theta_N)|
\label{eq:meanResultantLength}
\end{eqnarray}

\noindent The \emph{circular variance}, $CV$, is one minus the mean resultant length:

\begin{eqnarray}
CV(\theta_1, \ldots, \theta_N)=1-\bar{R}(\theta_1, \ldots, \theta_N)
\label{eq:cv}
\end{eqnarray}

\noindent The \emph{mean direction}, $\bar{\theta}$, is the angle of the resultant
vector:

\begin{eqnarray}
\bar{\theta}(\theta_1, \ldots, \theta_N)=\arg(\mathbf{R}(\theta_1, \ldots,
\theta_N))
\label{eq:meanDirection}
\end{eqnarray}

\noindent Note that the mean direction is not
defined when the resultant vector is zero, since the angle of the zero vector
is undefined.

\subsubsection{ITC}
\label{sec:itc}

The Inter-Trial Coherence (\gls{ITC}) is a measure of phase alignment among multiple
trials at a given time and frequency~\citep{tallonBaudryEtAl96,
delormeAndMakeig04}.  Mathematically, it is defined as the mean resultant
length ($\bar{R}$, Eq.~\ref{eq:meanResultantLength}) of the phases of the
trials at the given time and frequency.

\subsubsection{ERSP}
\label{sec:ersp}

The Event-Related Spectral Perturbation (ERSP) is the mean power of a set of
baseline-normalized epochs~\citep{delormeAndMakeig04}.

\subsubsection{Phase-amplitude coupling curve and modulation index}
\label{sec:pacCurveAndMI}

The \gls{PAC} curve is a graphical means of representing the
coupling between phases at a modulating frequency and amplitudes at a
modulated frequency. The \gls{MI} is a numerical quantification of the
strength of this coupling. We used the methods described in~\citet{tortEtAl10}
to compute both the \gls{PAC} curve and the \gls{MI}.

\subsection{Behavioral Data}
\label{sec:behavioralData}

\begin{figure}
\begin{center}
\includegraphics[width=2.5in]{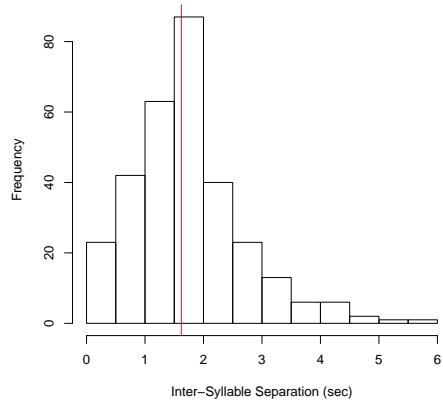}
\includegraphics[width=2.5in]{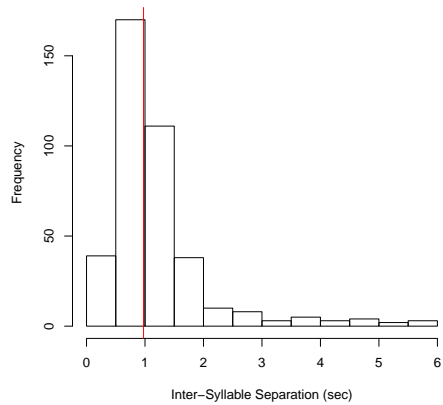}
\end{center}

\caption{Histogram of inter-syllable separations from the experimental
sessions with the largest (1.62~sec) and shortest (0.97~sec) intersyllable
separation time, EC2\_B105 (a) and EC2\_B89 (b), respectively.}

\label{fig:issHist}
\end{figure}

Subjects produced consonant-vowel syllables at different speeds in different
experimental sessions. This manuscript characterizes data in the experimental
sessions with largest and shortest median intersyllable separation times
(Figure~\ref{fig:issHist}).

\subsection{ITC across the vSMC}
\label{sec:itcsAcrossvSMC}

\begin{figure}
\begin{center}
\includegraphics[width=3.5in]{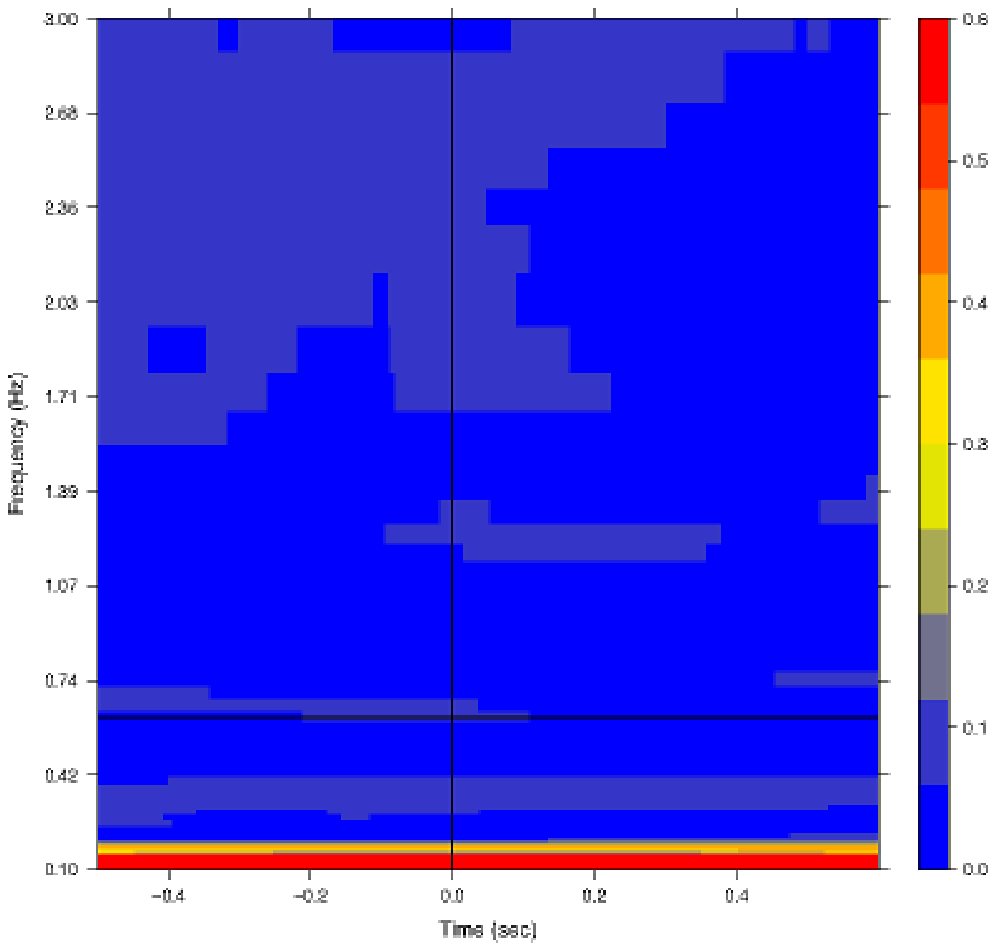}
\end{center}

\caption{ITC for electrode 129 computed from recordings in experimental
session EC2\_B105.}

\label{fig:itc129B105}
\end{figure}

\begin{figure}
\begin{center}
\includegraphics[width=3.5in]{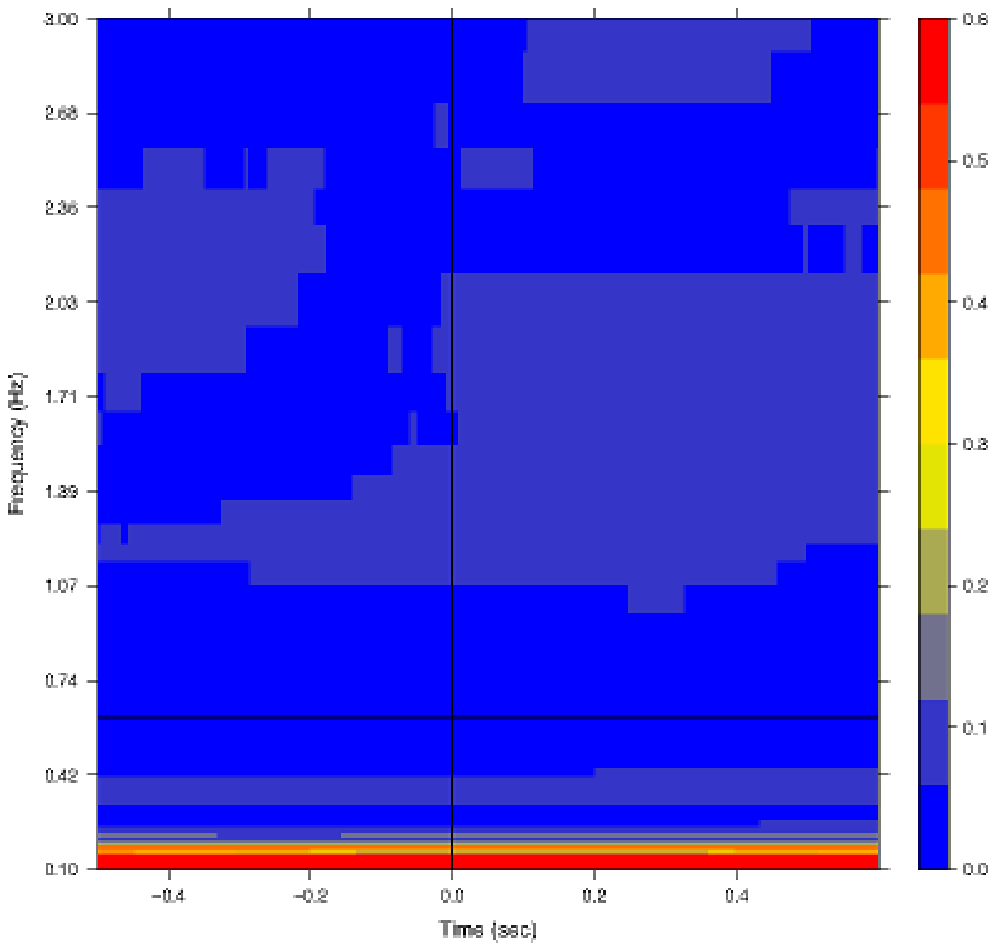}
\end{center}

\caption{ITC for electrode 130 computed from recordings in experimental
session EC2\_B105.}

\label{fig:itc130B105}
\end{figure}

\begin{figure}
\begin{center}
\includegraphics[width=3.5in]{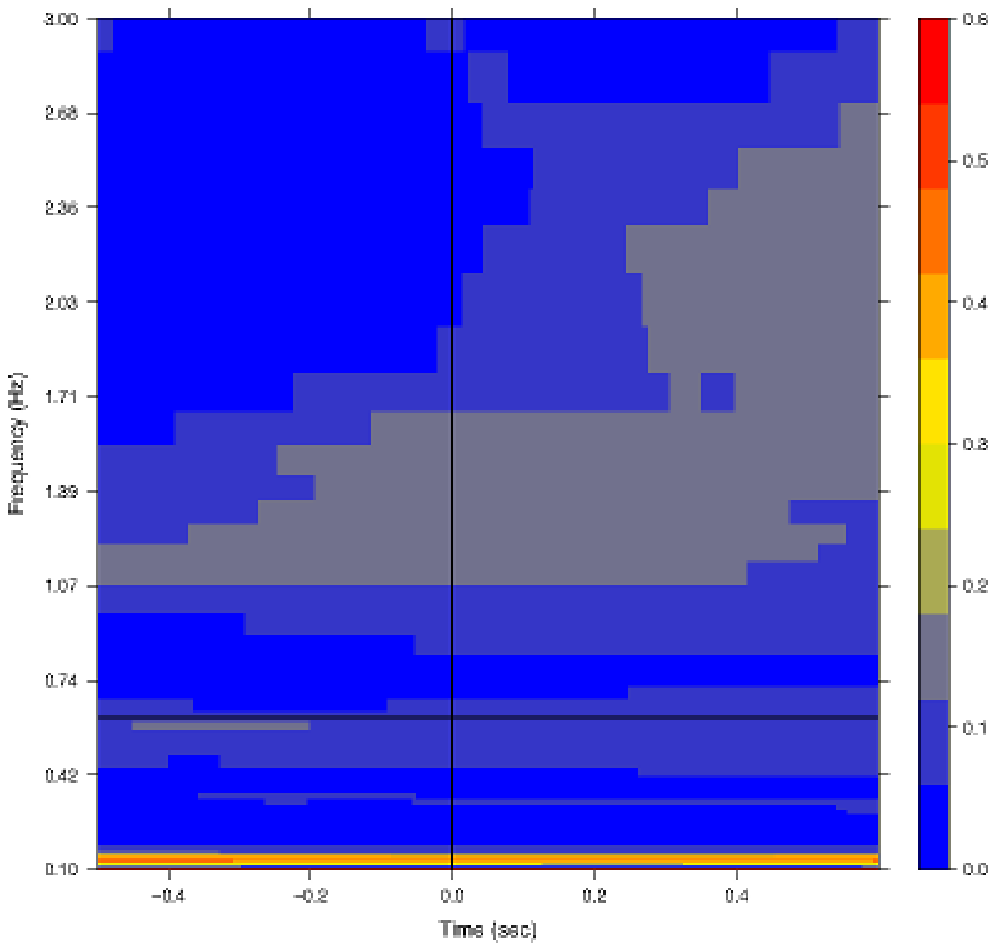}
\end{center}

\caption{ITC for electrode 131 computed from recordings in experimental
session EC2\_B105.}

\label{fig:itc131B105}
\end{figure}

\begin{figure}
\begin{center}
\includegraphics[width=3.5in]{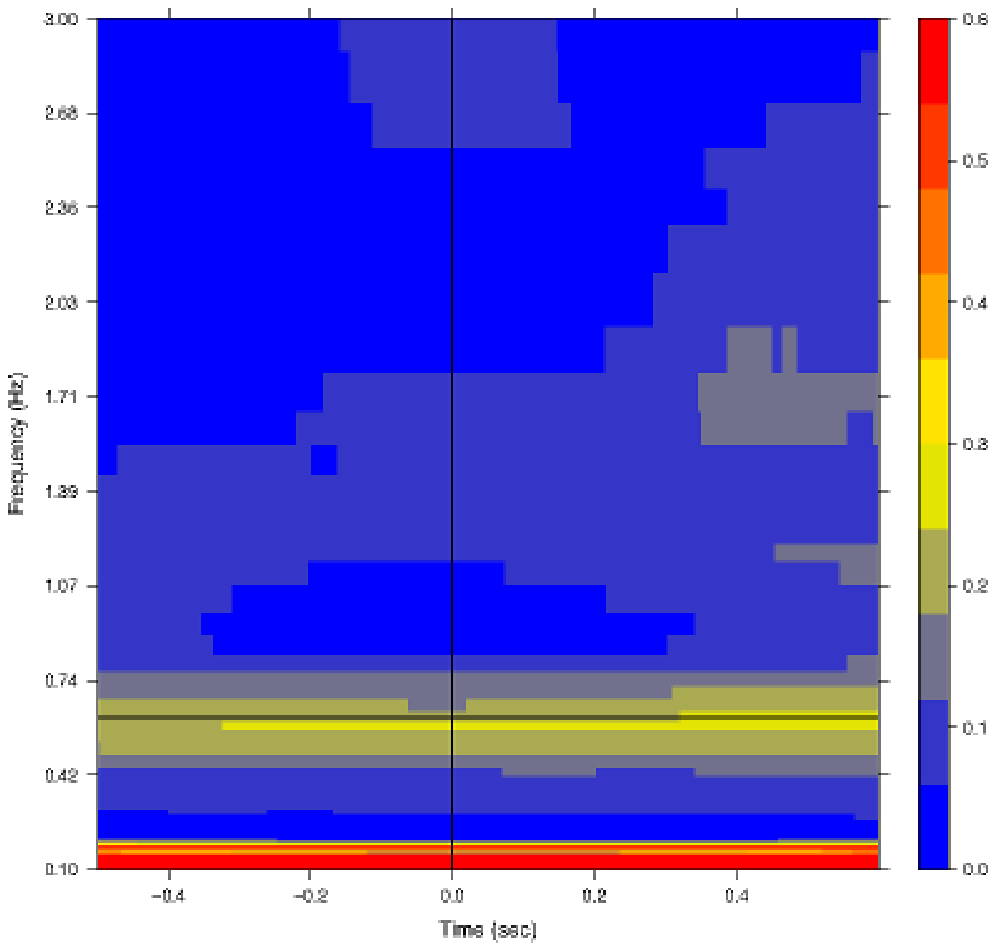}
\end{center}

\caption{ITC for electrode 132 computed from recordings in experimental
session EC2\_B105.}

\label{fig:itc132B105}
\end{figure}

\begin{figure}
\begin{center}
\includegraphics[width=3.5in]{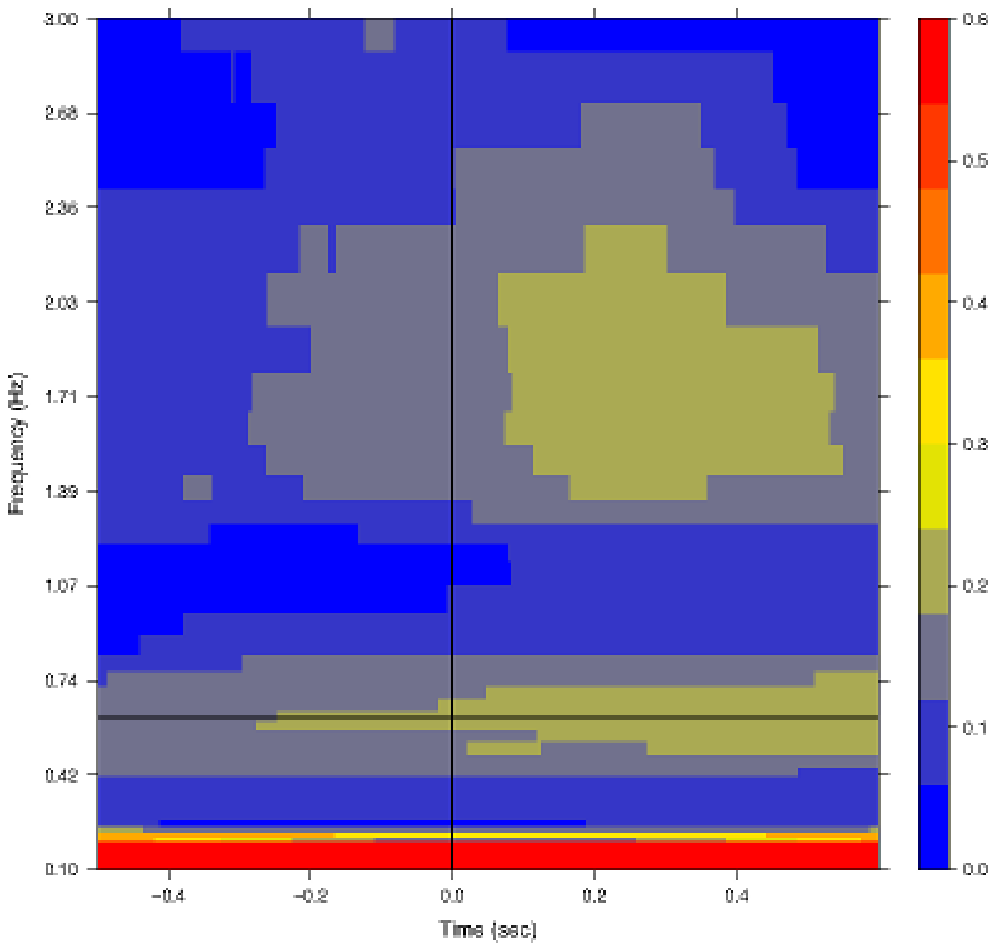}
\end{center}

\caption{ITC for electrode 133 computed from recordings in experimental
session EC2\_B105.}

\label{fig:itc133B105}
\end{figure}

\begin{figure}
\begin{center}
\includegraphics[width=3.5in]{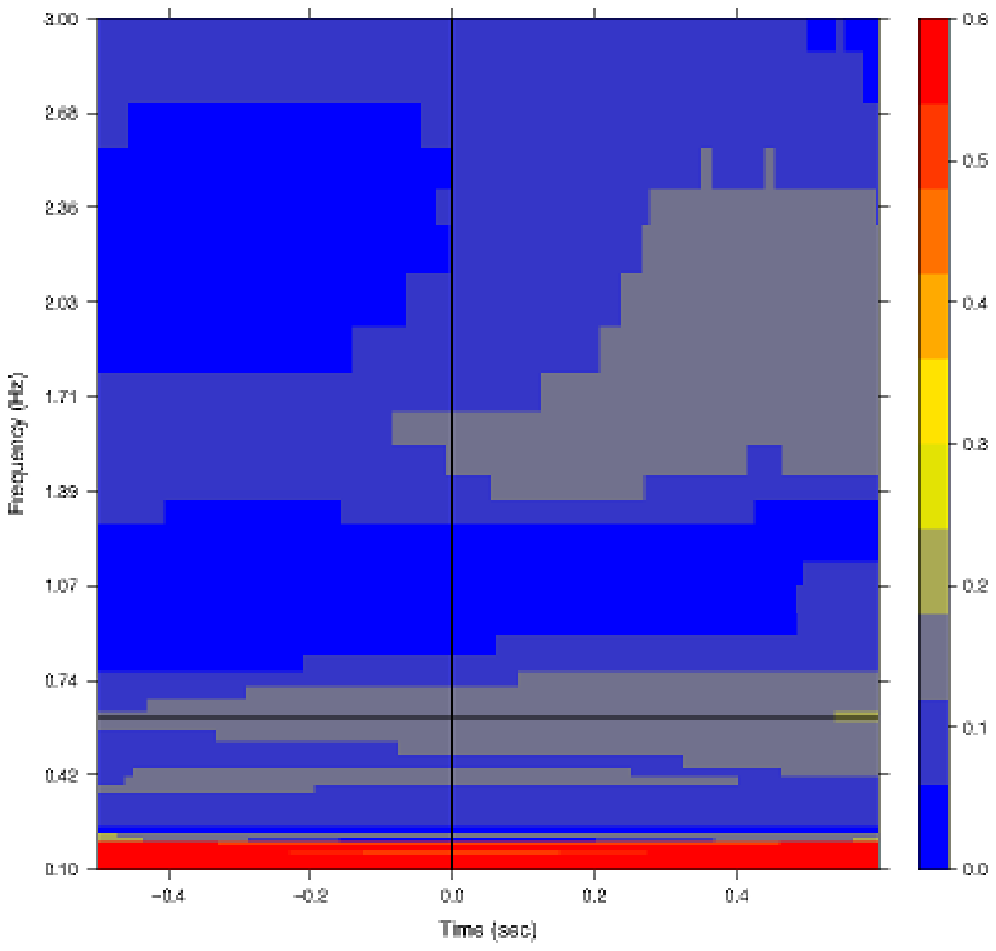}
\end{center}

\caption{ITC for electrode 134 computed from recordings in experimental
session EC2\_B105.}

\label{fig:itc134B105}
\end{figure}

\begin{figure}
\begin{center}
\includegraphics[width=3.5in]{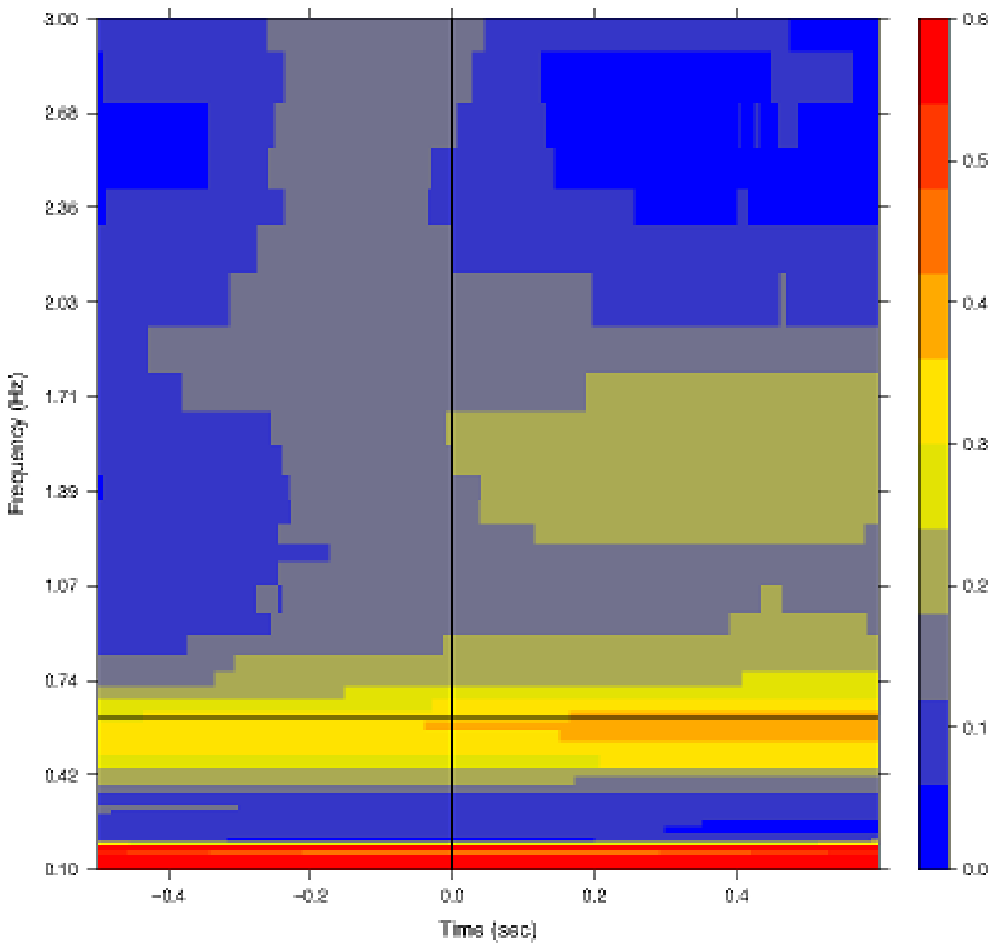}
\end{center}

\caption{ITC for electrode 135 computed from recordings in experimental
session EC2\_B105.}

\label{fig:itc135B105}
\end{figure}

\begin{figure}
\begin{center}
\includegraphics[width=3.5in]{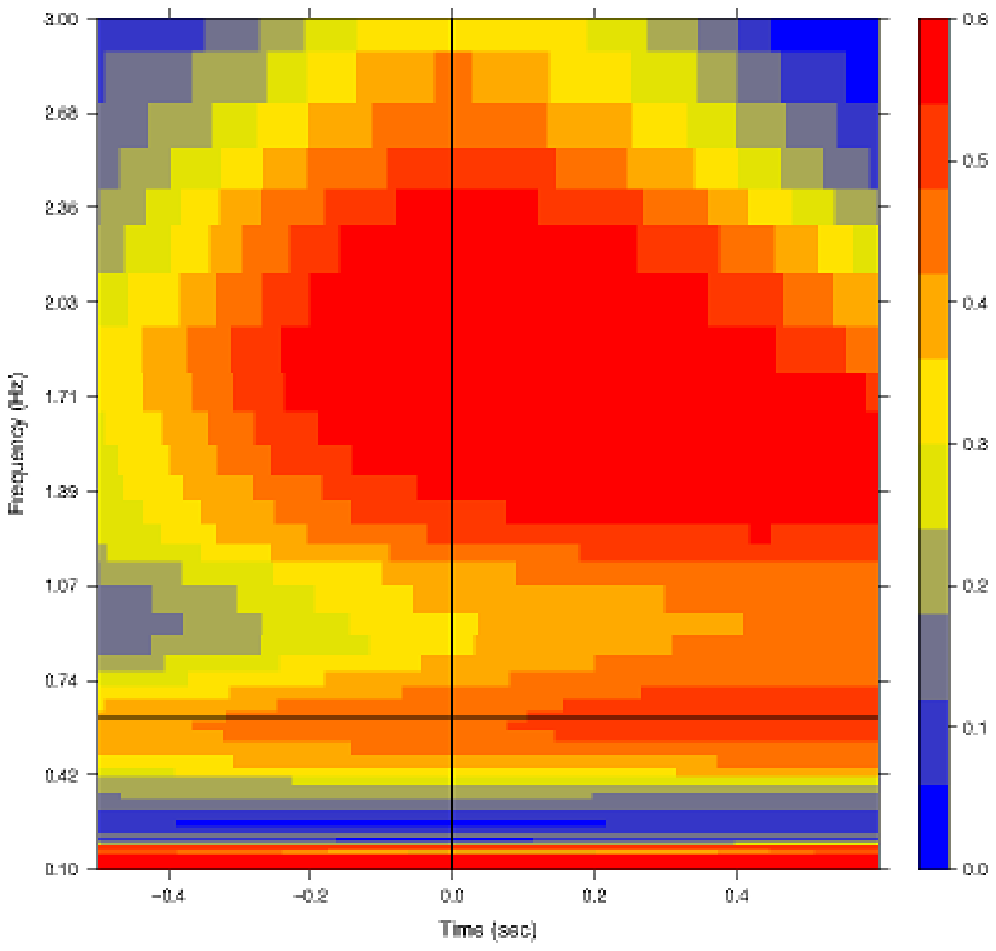}
\end{center}

\caption{ITC for electrode 136 computed from recordings in experimental
session EC2\_B105.}

\label{fig:itc136B105}
\end{figure}

\begin{figure}
\begin{center}
\includegraphics[width=3.5in]{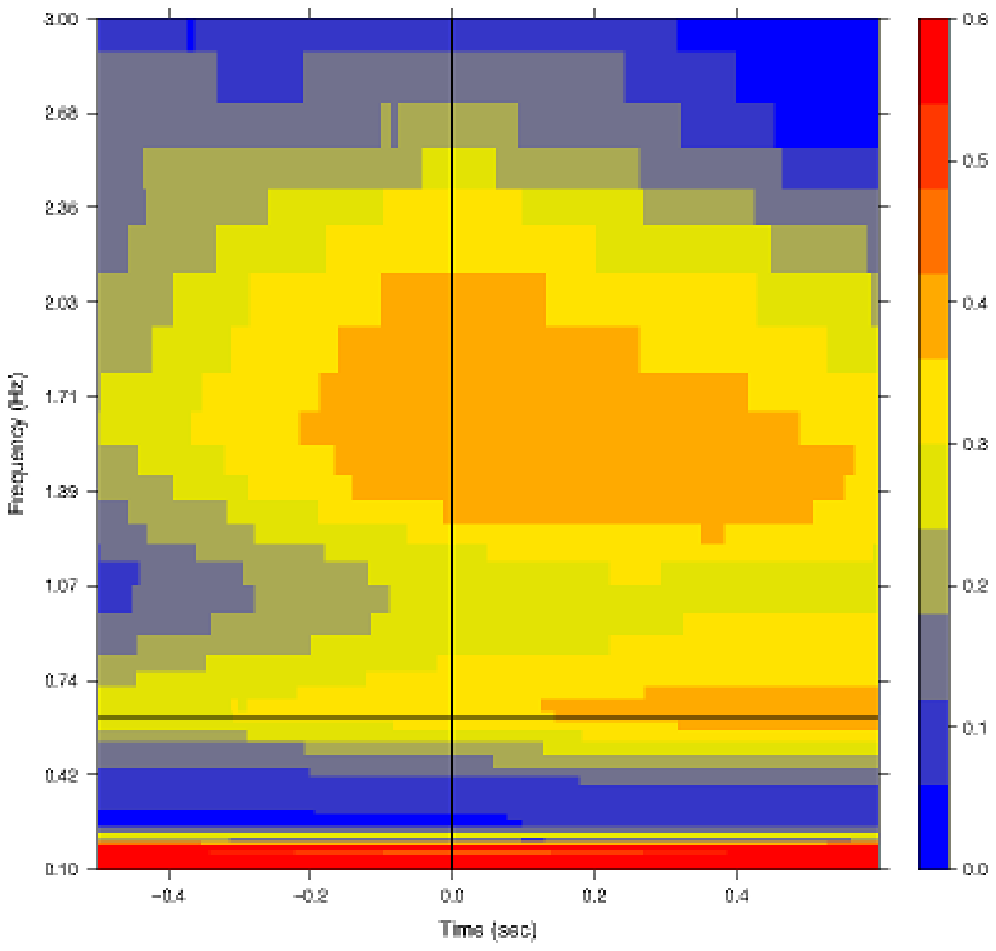}
\end{center}

\caption{ITC for electrode 137 computed from recordings in experimental
session EC2\_B105.}

\label{fig:itc137B105}
\end{figure}

\begin{figure}
\begin{center}
\includegraphics[width=3.5in]{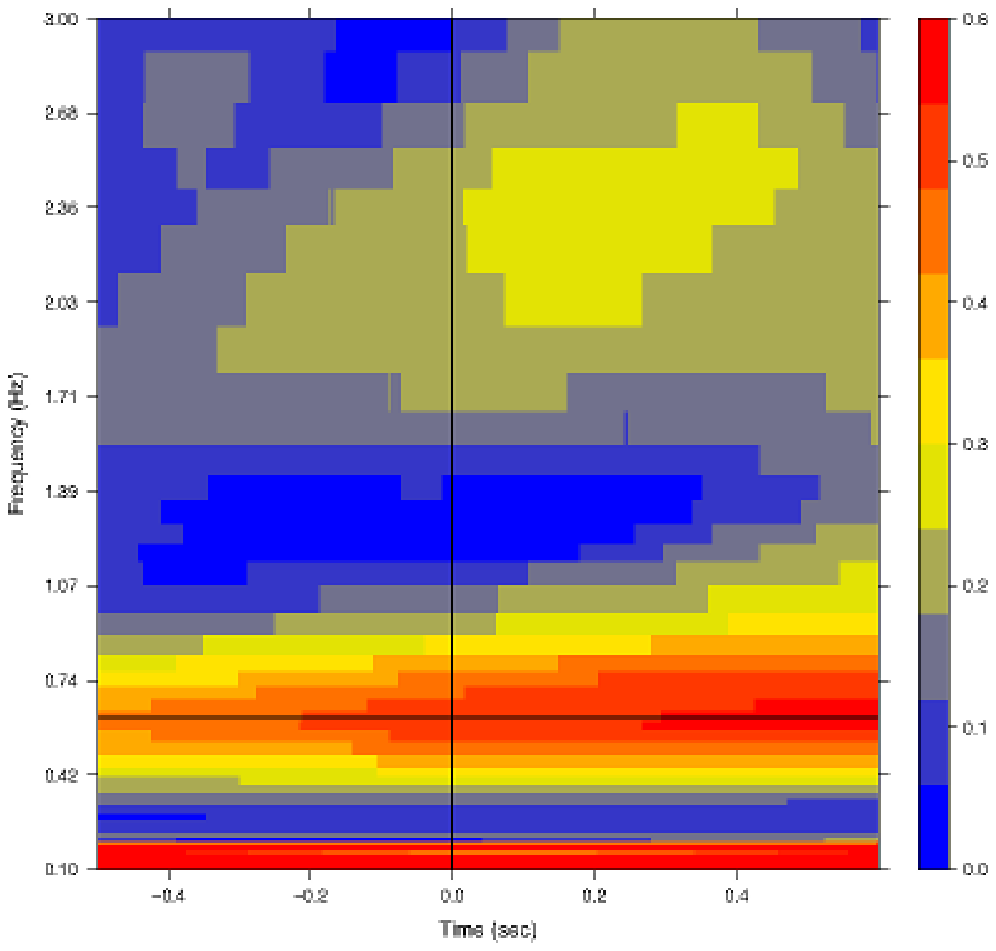}
\end{center}

\caption{ITC for electrode 138 computed from recordings in experimental
session EC2\_B105.}

\label{fig:itc138B105}
\end{figure}

\begin{figure}
\begin{center}
\includegraphics[width=3.5in]{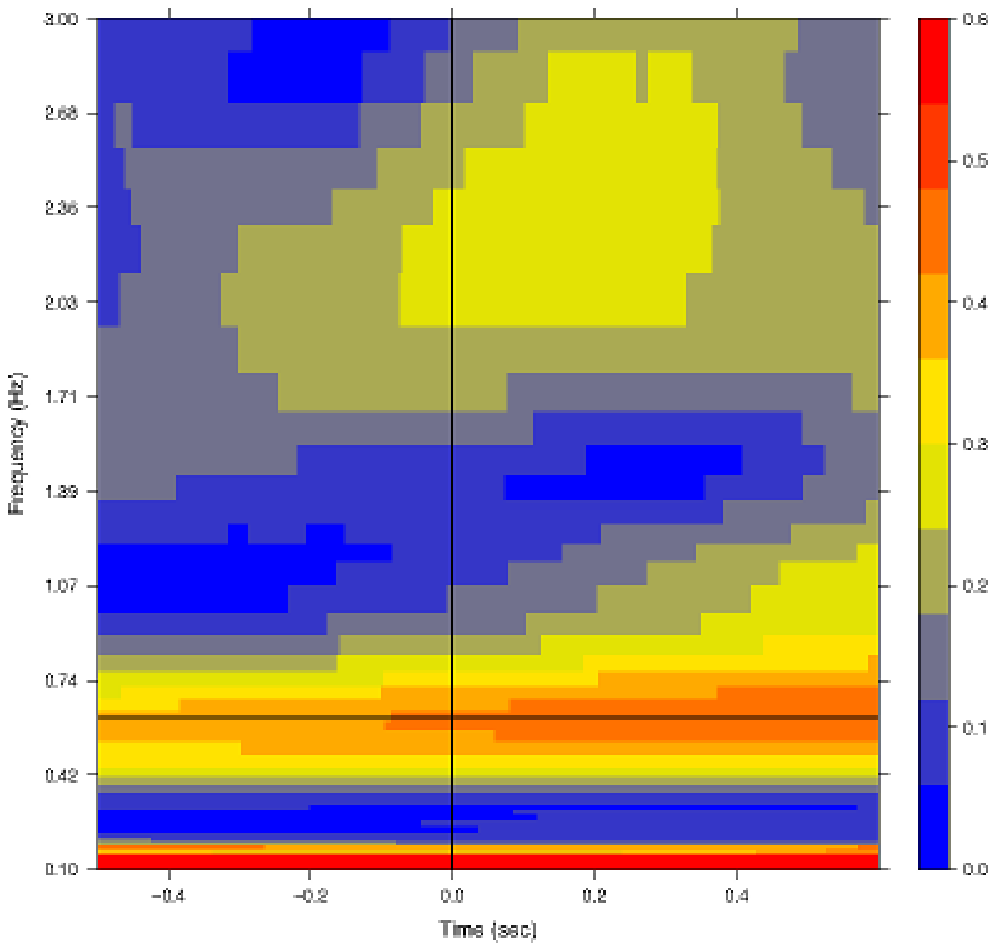}
\end{center}

\caption{ITC for electrode 139 computed from recordings in experimental
session EC2\_B105.}

\label{fig:itc139B105}
\end{figure}

\begin{figure}
\begin{center}
\includegraphics[width=3.5in]{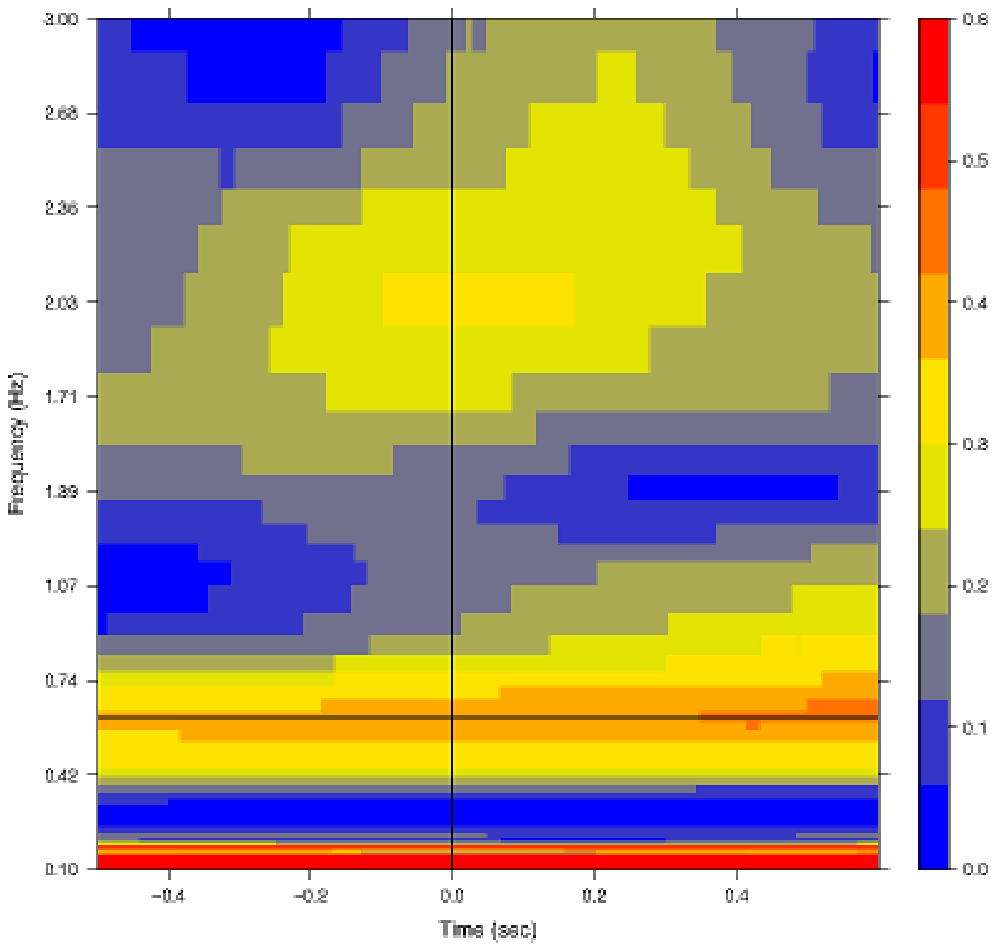}
\end{center}

\caption{ITC for electrode 140 computed from recordings in experimental
session EC2\_B105.}

\label{fig:itc140B105}
\end{figure}

\begin{figure}
\begin{center}
\includegraphics[width=3.5in]{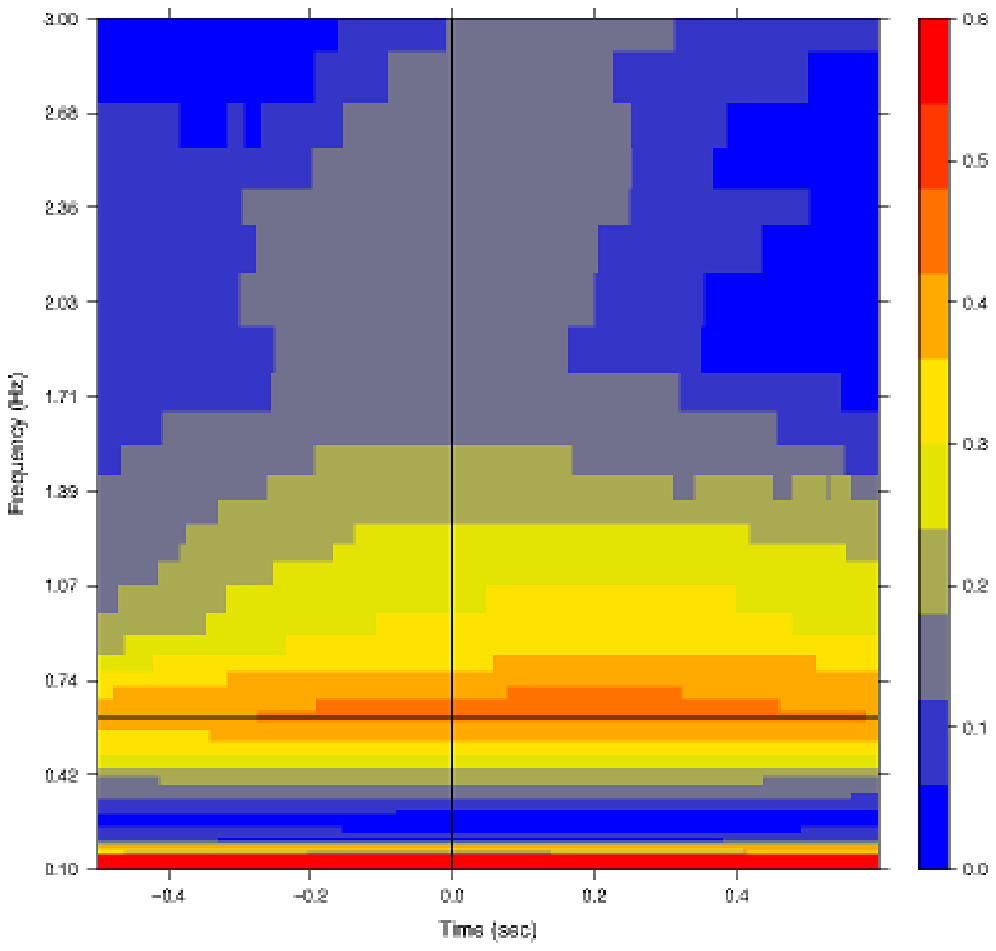}
\end{center}

\caption{ITC for electrode 141 computed from recordings in experimental
session EC2\_B105.}

\label{fig:itc141B105}
\end{figure}

\begin{figure}
\begin{center}
\includegraphics[width=3.5in]{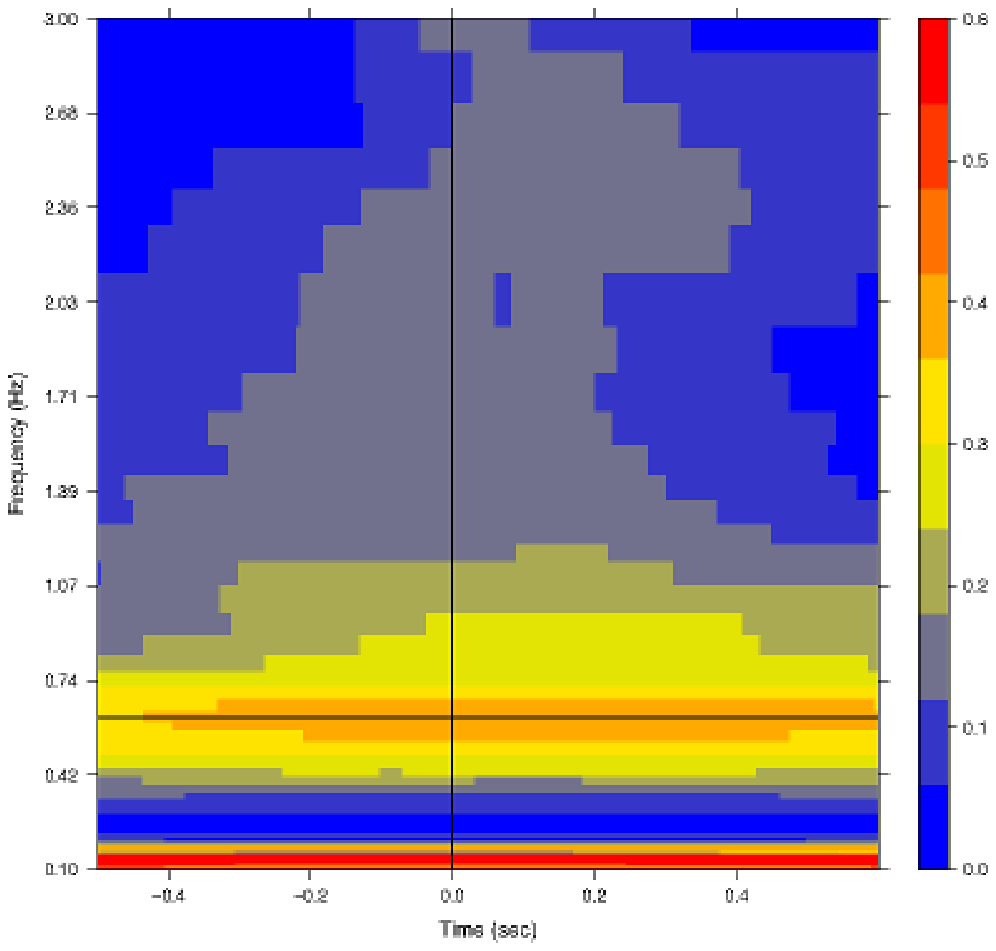}
\end{center}

\caption{ITC for electrode 142 computed from recordings in experimental
session EC2\_B105.}

\label{fig:itc142B105}
\end{figure}

\begin{figure}
\begin{center}
\includegraphics[width=3.5in]{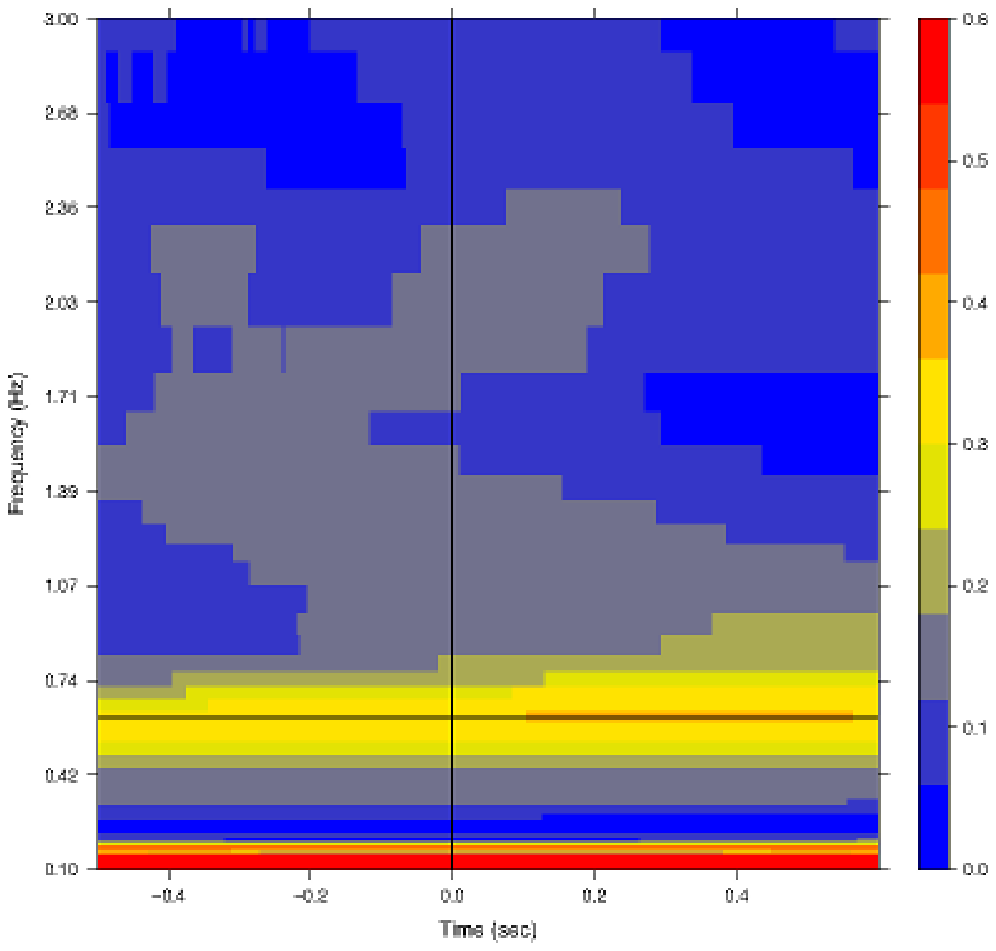}
\end{center}

\caption{ITC for electrode 143 computed from recordings in experimental
session EC2\_B105.}

\label{fig:itc143B105}
\end{figure}

\begin{figure}
\begin{center}
\includegraphics[width=3.5in]{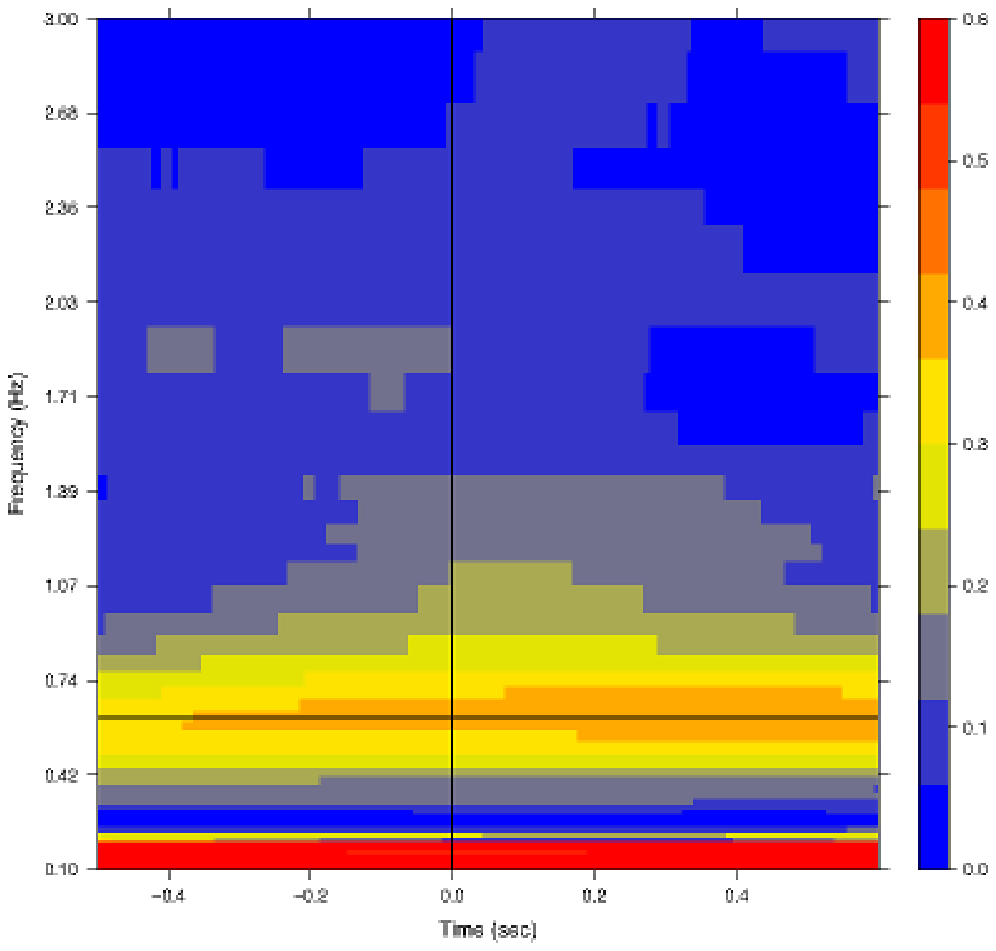}
\end{center}

\caption{ITC for electrode 144 computed from recordings in experimental
session EC2\_B105.}

\label{fig:itc144B105}
\end{figure}

Largest coherence values at the entrained frequency were observed over the
\gls{vSMC}. Figures~\ref{fig:itc129B105}-\ref{fig:itc144B105} plot the \gls{ITC}, computed
from recordings in experimental session EC2\_B105, between the ventral
electrode 129 and the dorsal electrode 144. For electrodes over the temporal
cortex there is almost no \gls{ITC}
(Figures~\ref{fig:itc129B105}-\ref{fig:itc131B105}). \gls{ITC} increases over the
ventral section of the \gls{vSMC}
(Figures~\ref{fig:itc132B105}-\ref{fig:itc135B105}), peaks at its center
(Figures~\ref{fig:itc136B105}-\ref{fig:itc139B105}), and tapers over its dorsal
section (Figures~\ref{fig:itc140B105}-\ref{fig:itc144B105}).

\clearpage

\subsection{ERSPs across the vSMC}
\label{sec:erspsAcrossvSMC}

\begin{figure}
\begin{center}
\includegraphics[width=3.5in]{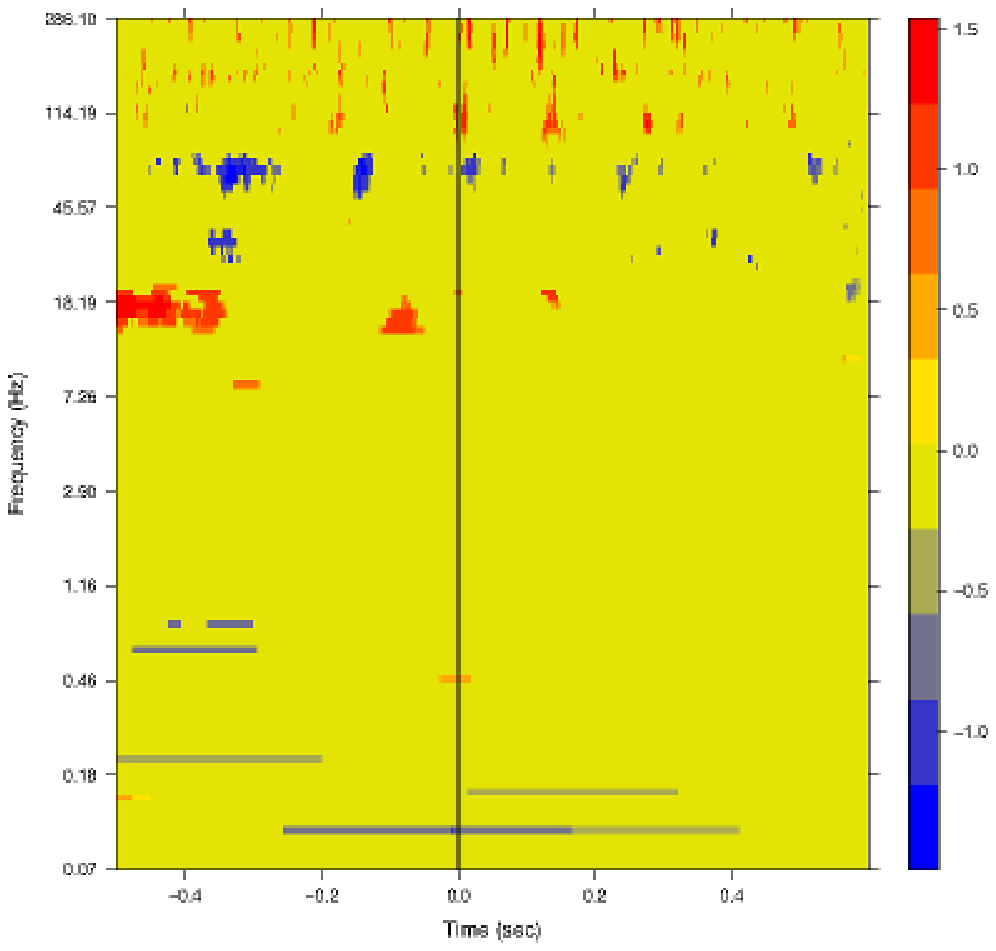}
\end{center}

\caption{ERSP for electrode 129 computed from recordings in experimental
session EC2\_B105.}

\label{fig:ersp129B105}
\end{figure}

\begin{figure}
\begin{center}
\includegraphics[width=3.5in]{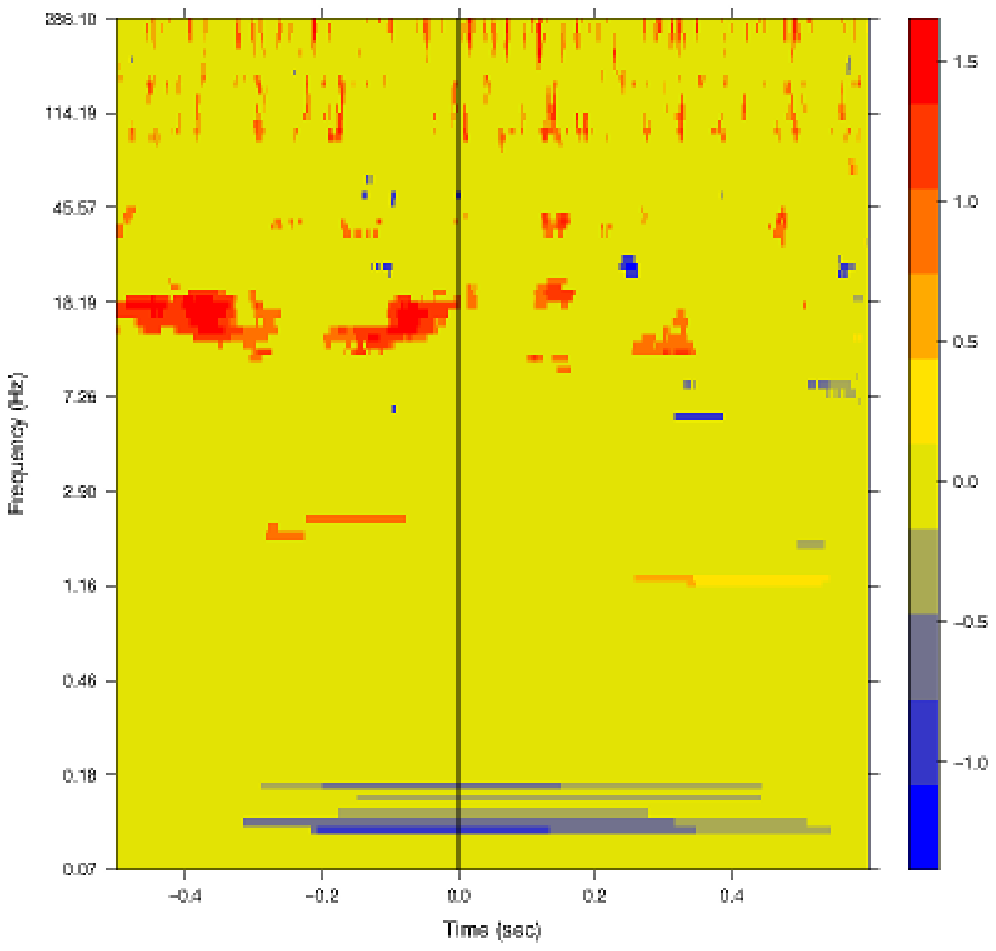}
\end{center}

\caption{ERSP for electrode 130 computed from recordings in experimental
session EC2\_B105.}

\label{fig:ersp130B105}
\end{figure}

\begin{figure}
\begin{center}
\includegraphics[width=3.5in]{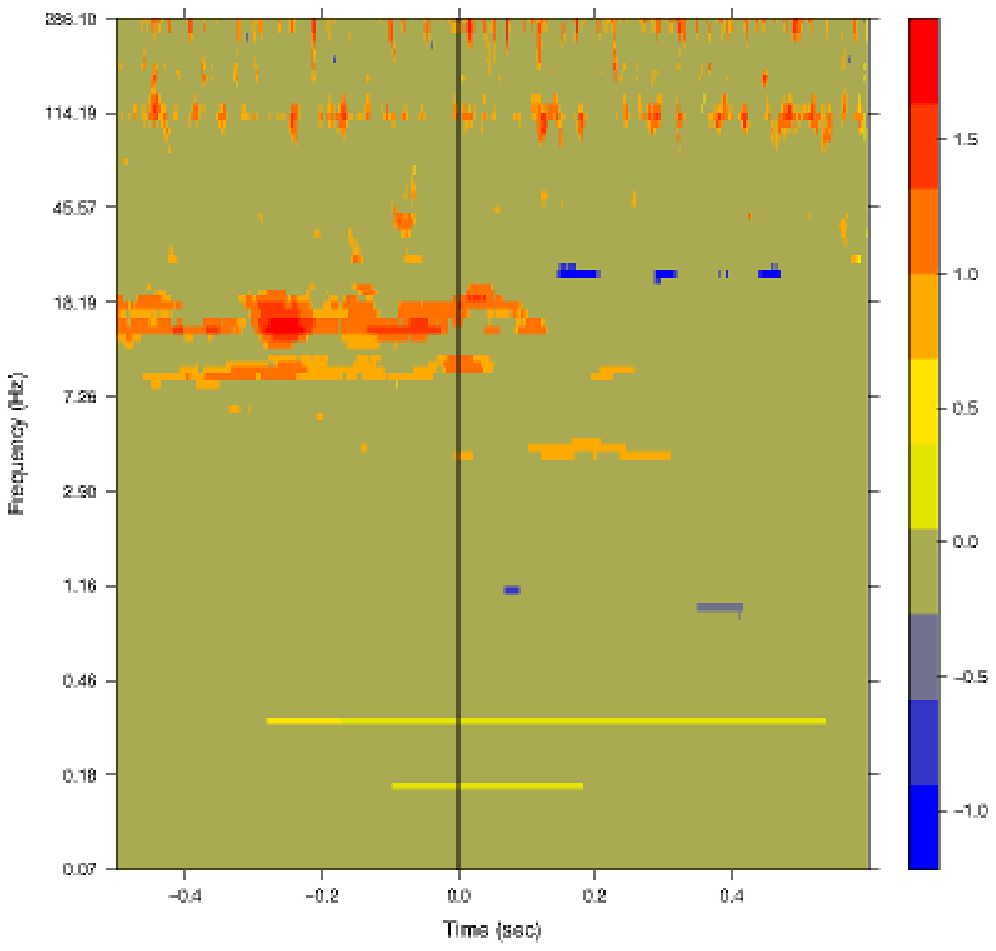}
\end{center}

\caption{ERSP for electrode 131 computed from recordings in experimental
session EC2\_B105.}

\label{fig:ersp131B105}
\end{figure}

\begin{figure}
\begin{center}
\includegraphics[width=3.5in]{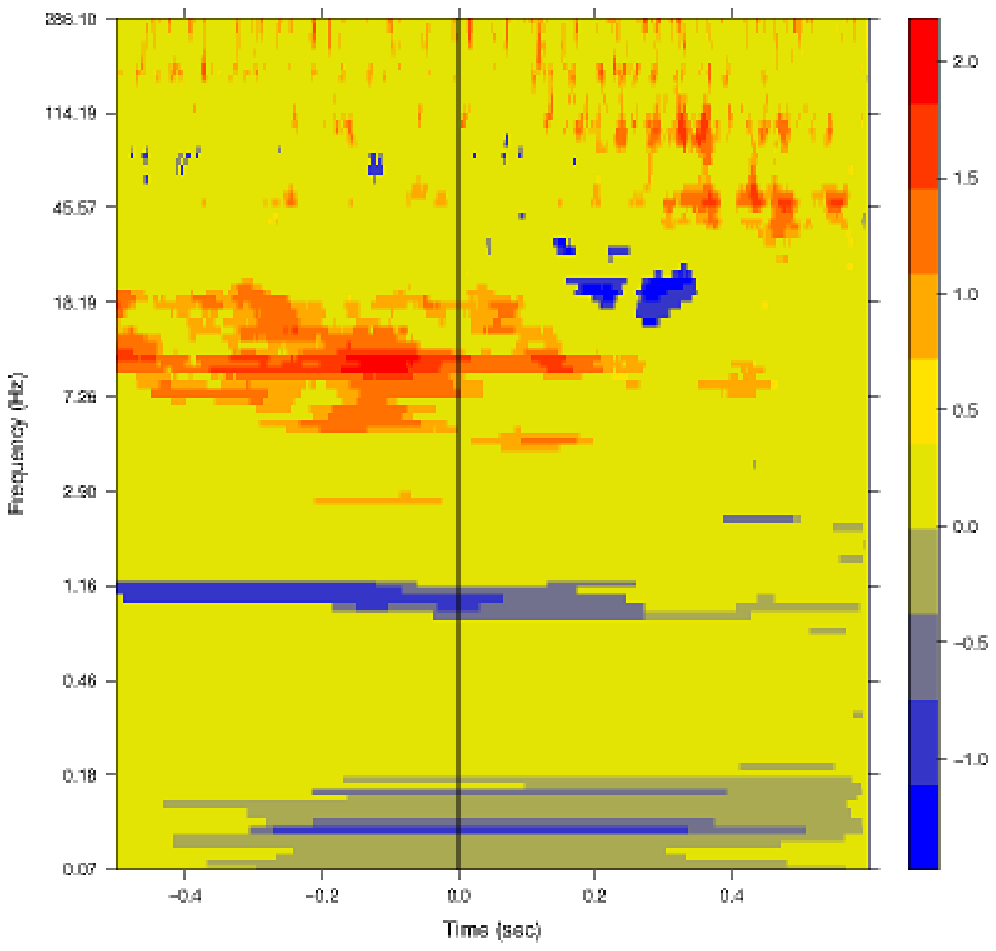}
\end{center}

\caption{ERSP for electrode 132 computed from recordings in experimental
session EC2\_B105.}

\label{fig:ersp132B105}
\end{figure}

\begin{figure}
\begin{center}
\includegraphics[width=3.5in]{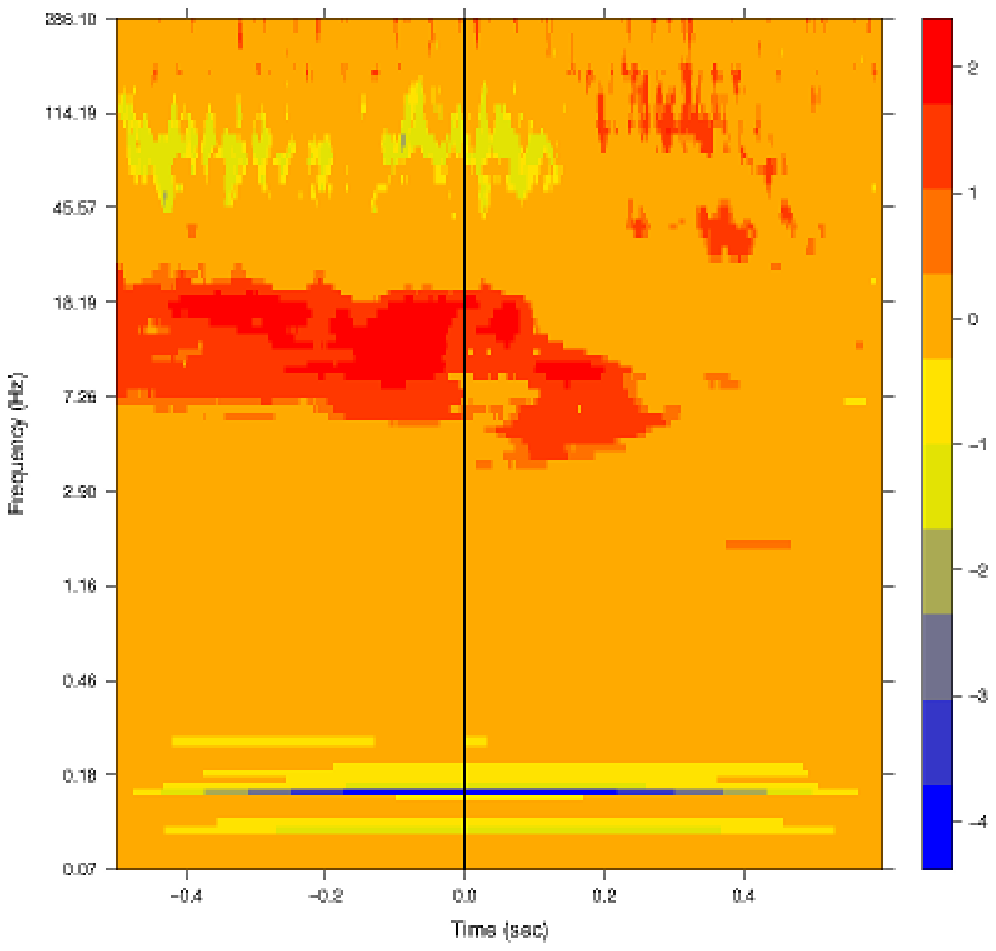}
\end{center}

\caption{ERSP for electrode 133 computed from recordings in experimental
session EC2\_B105.}

\label{fig:ersp133B105}
\end{figure}

\begin{figure}
\begin{center}
\includegraphics[width=3.5in]{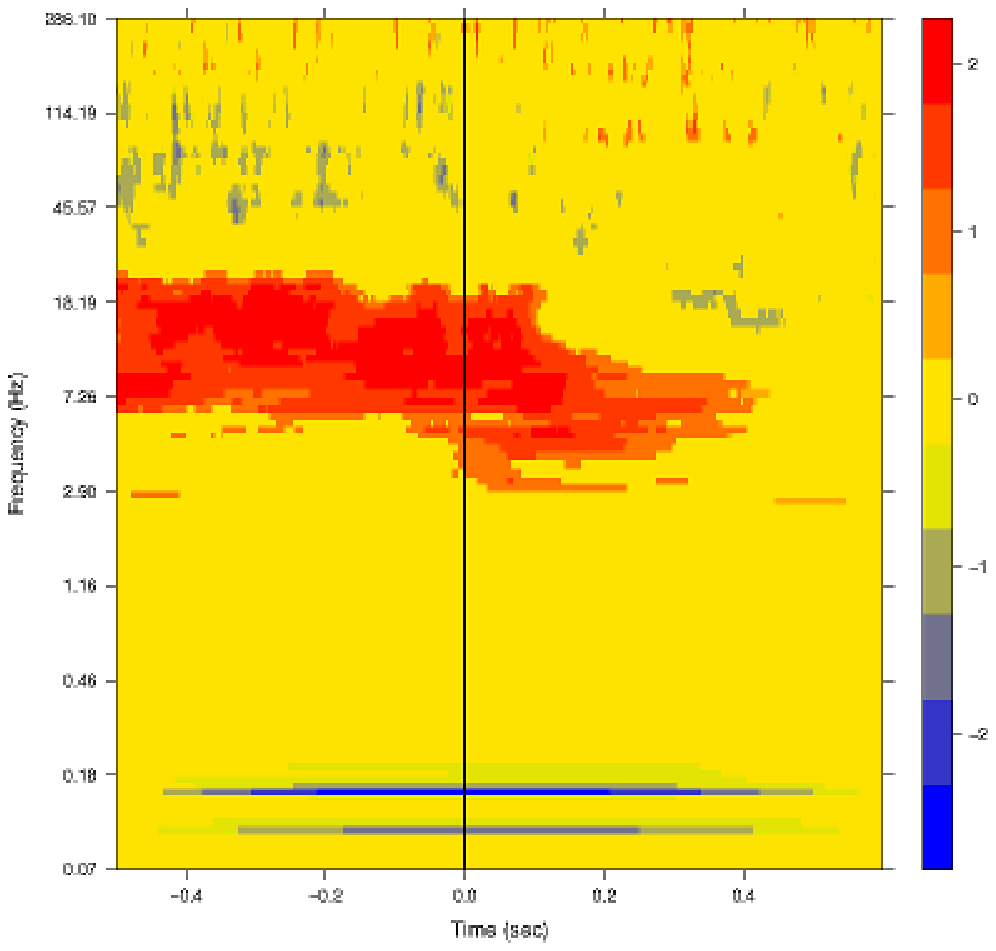}
\end{center}

\caption{ERSP for electrode 134 computed from recordings in experimental
session EC2\_B105.}

\label{fig:ersp134B105}
\end{figure}

\begin{figure}
\begin{center}
\includegraphics[width=3.5in]{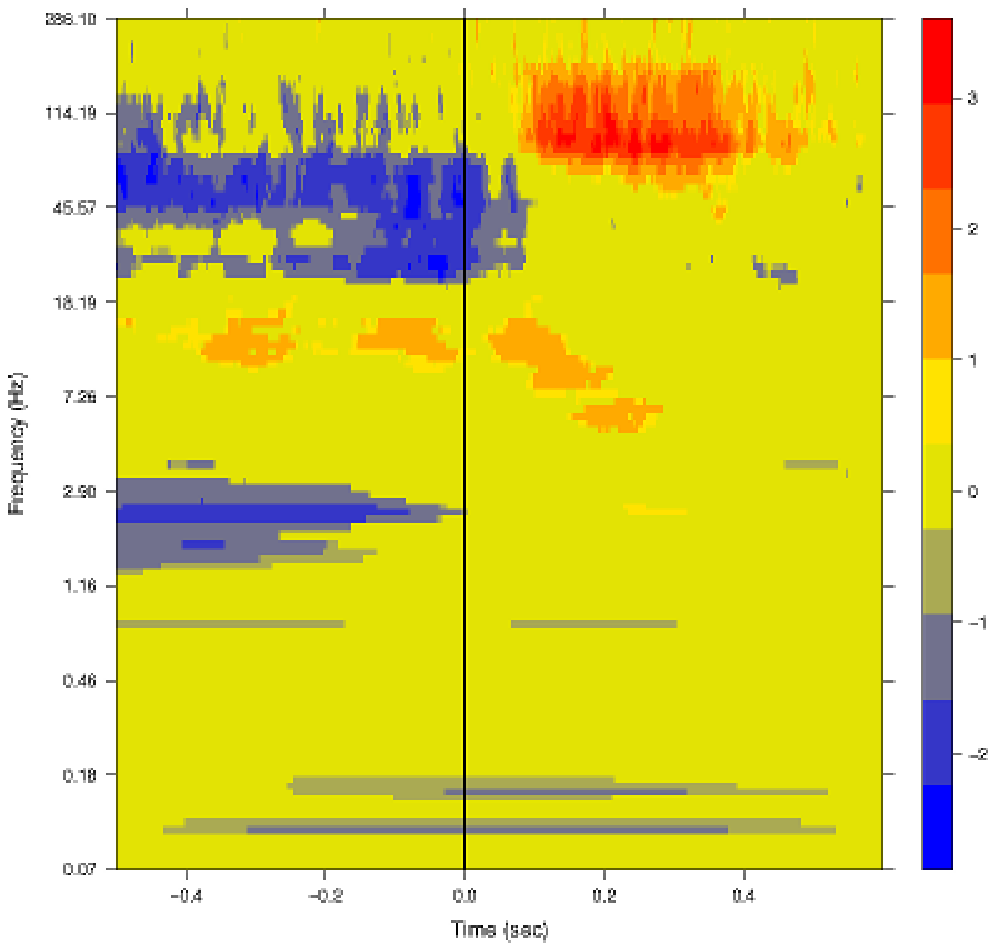}
\end{center}

\caption{ERSP for electrode 135 computed from recordings in experimental
session EC2\_B105.}

\label{fig:ersp135B105}
\end{figure}

\begin{figure}
\begin{center}
\includegraphics[width=3.5in]{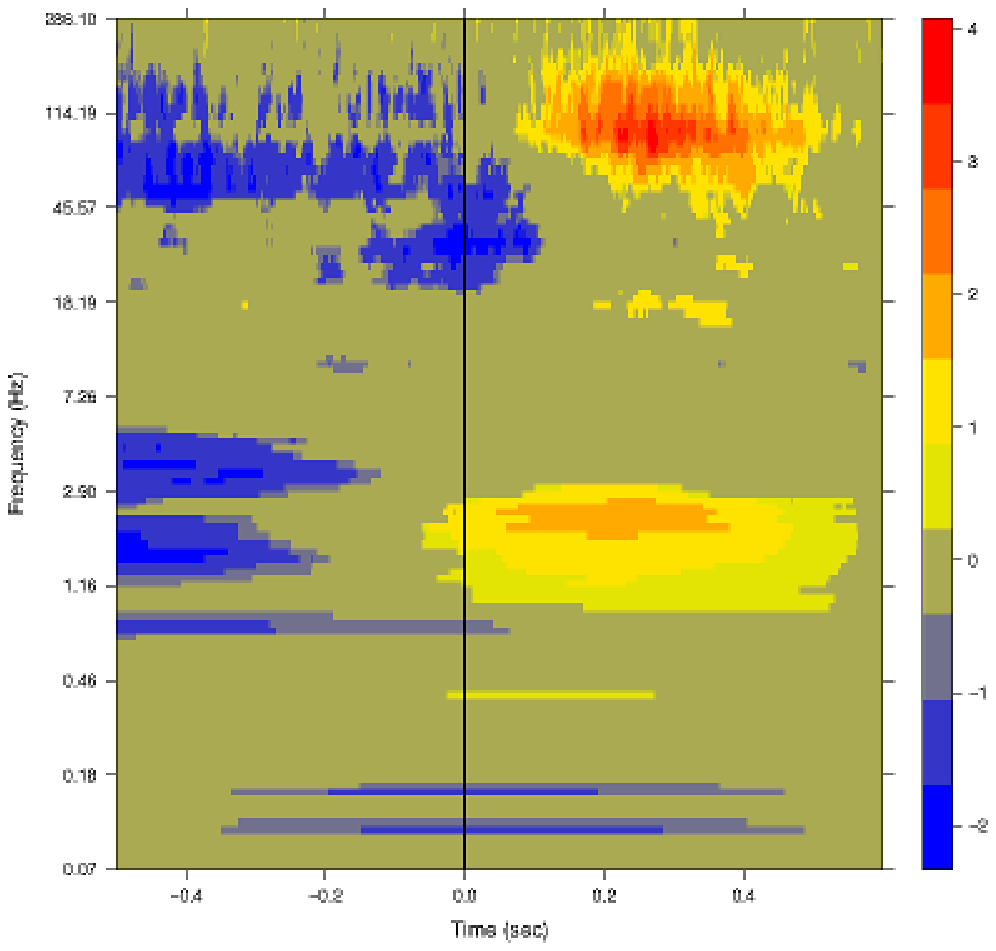}
\end{center}

\caption{ERSP for electrode 136 computed from recordings in experimental
session EC2\_B105.}

\label{fig:ersp136B105}
\end{figure}

\begin{figure}
\begin{center}
\includegraphics[width=3.5in]{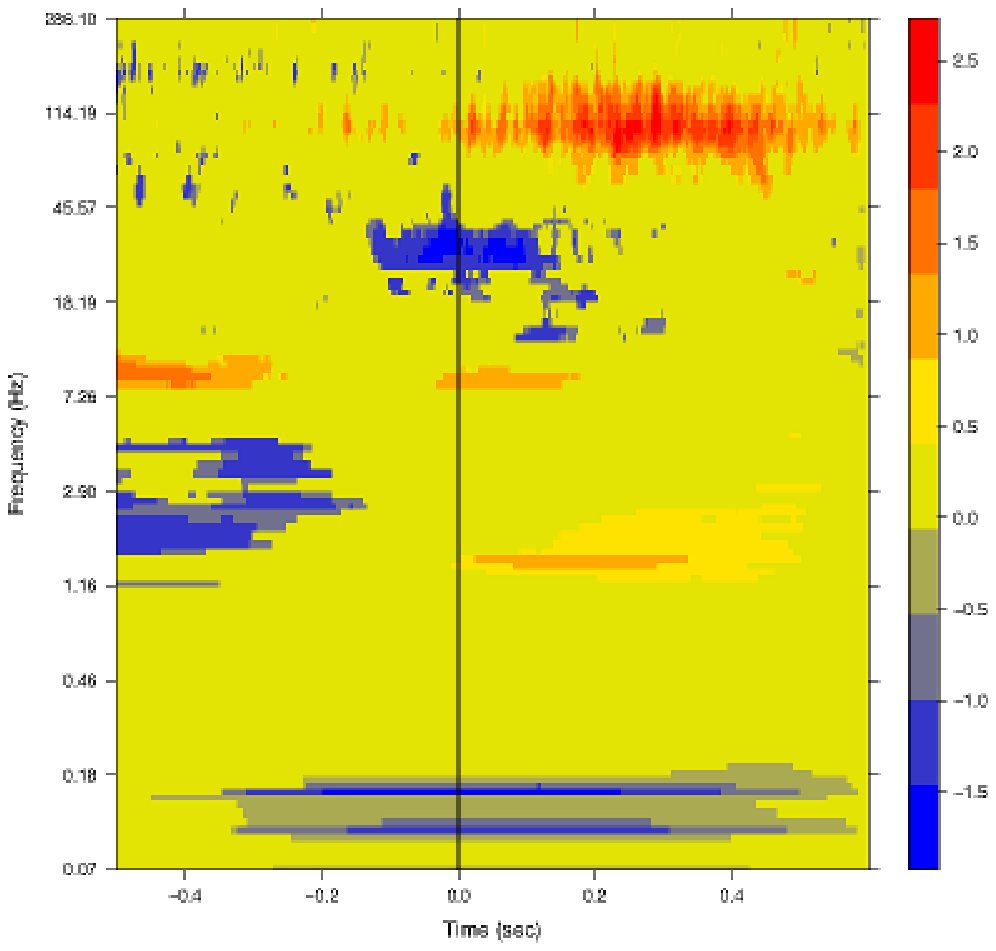}
\end{center}

\caption{ERSP for electrode 137 computed from recordings in experimental
session EC2\_B105.}

\label{fig:ersp137B105}
\end{figure}

\begin{figure}
\begin{center}
\includegraphics[width=3.5in]{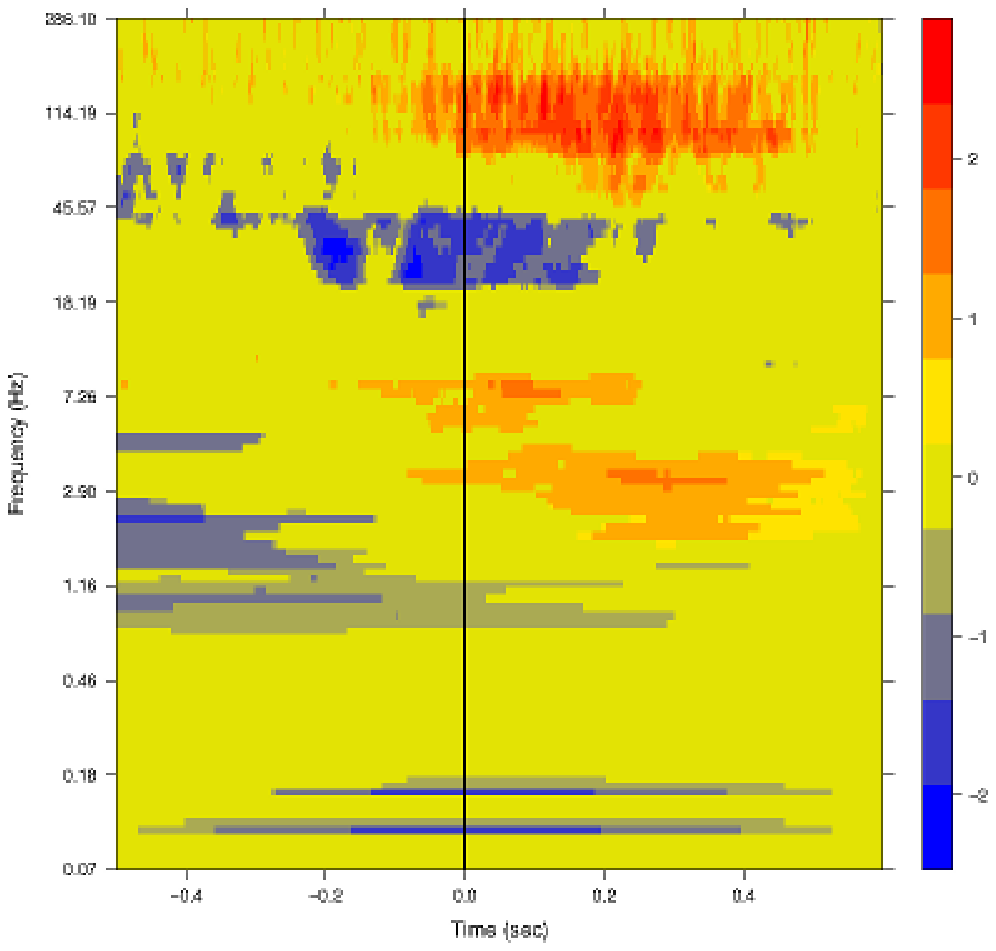}
\end{center}

\caption{ERSP for electrode 138 computed from recordings in experimental
session EC2\_B105.}

\label{fig:ersp138B105}
\end{figure}

\begin{figure}
\begin{center}
\includegraphics[width=3.5in]{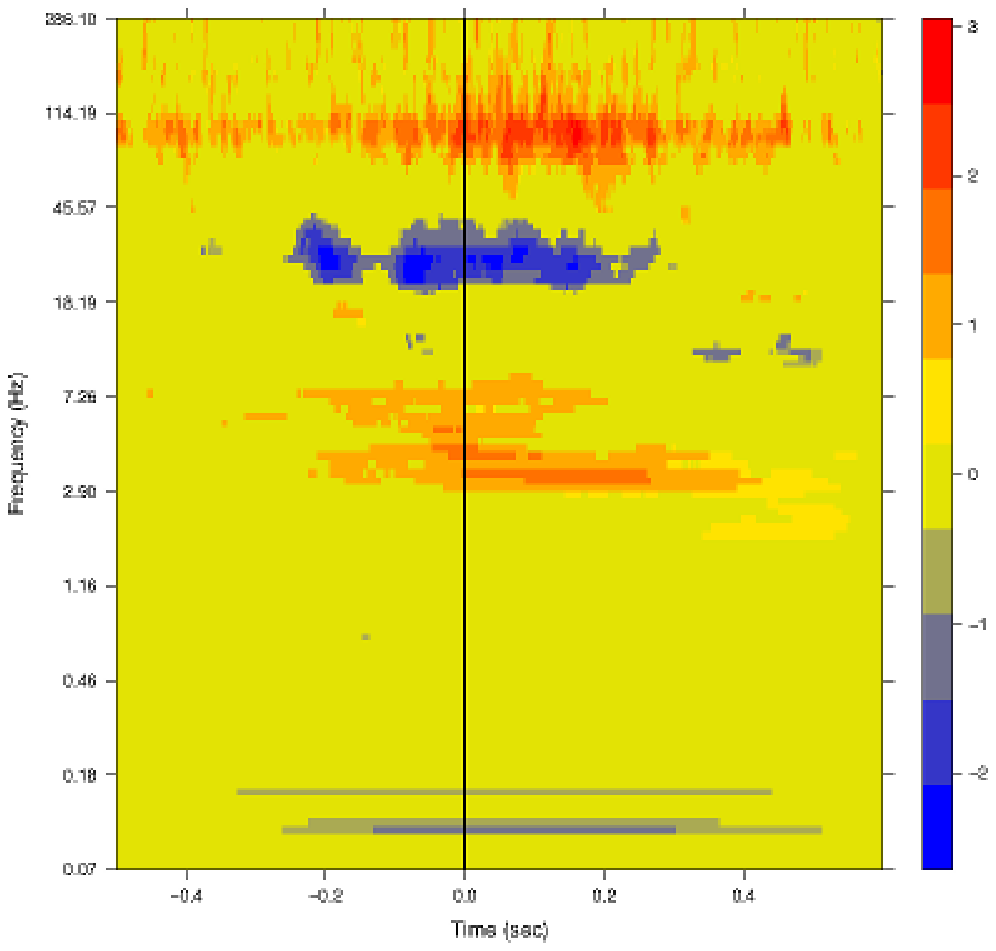}
\end{center}

\caption{ERSP for electrode 139 computed from recordings in experimental
session EC2\_B105.}

\label{fig:ersp139B105}
\end{figure}

\begin{figure}
\begin{center}
\includegraphics[width=3.5in]{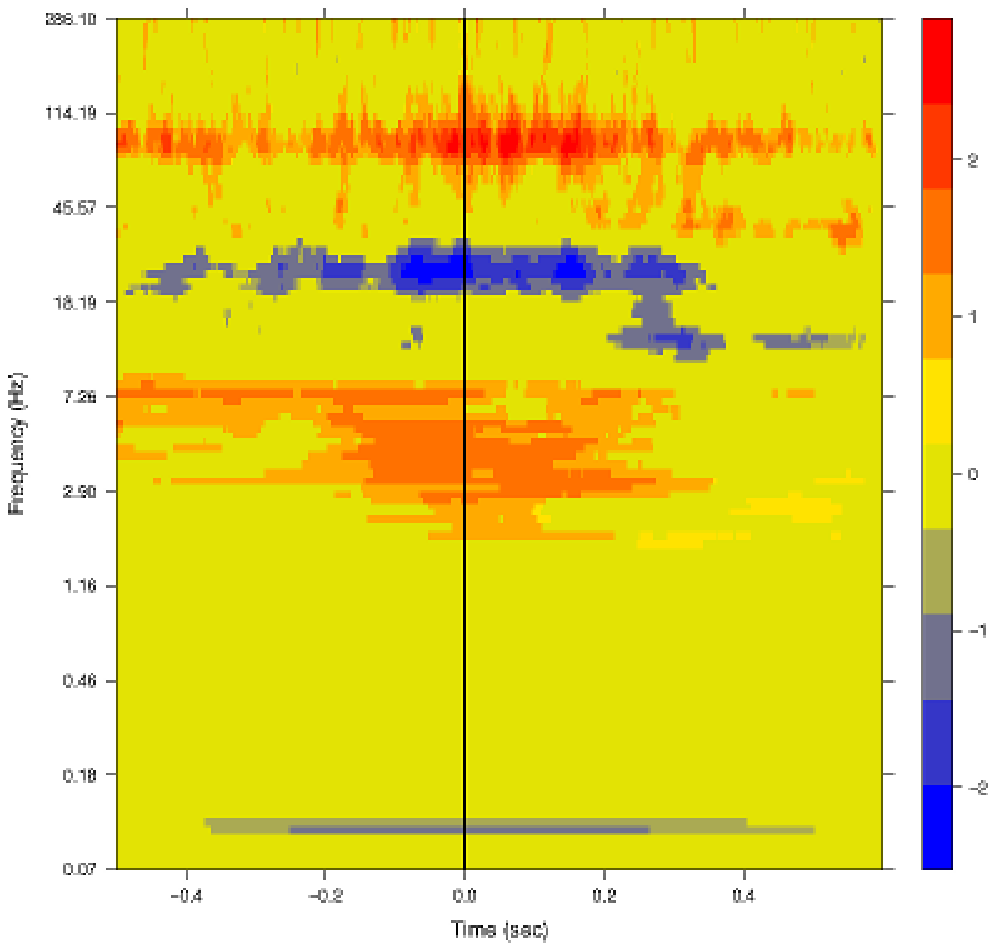}
\end{center}

\caption{ERSP for electrode 140 computed from recordings in experimental
session EC2\_B105.}

\label{fig:ersp140B105}
\end{figure}

\begin{figure}
\begin{center}
\includegraphics[width=3.5in]{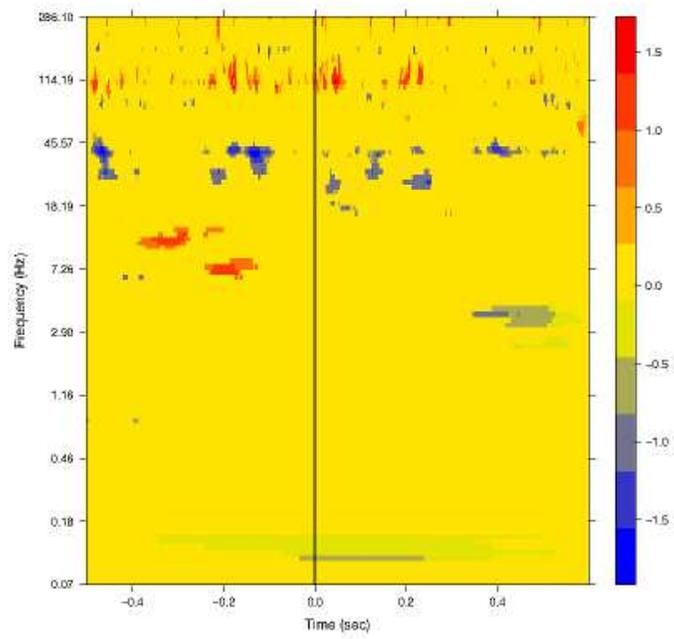}
\end{center}

\caption{ERSP for electrode 141 computed from recordings in experimental
session EC2\_B105.}

\label{fig:ersp141B105}
\end{figure}

Figures~\ref{fig:ersp129B105}-\ref{fig:ersp141B105} plot \glspl{ERSP} at
electrodes along the ventro-dorsal axis of the \gls{vSMC}. Note the sharp
transition between the peak of large evoked beta power before the \gls{CV}
transition at electrodes 132-135 over the auditory cortex to the peak of large
evoked high-gamma power after the \gls{CV} transition at electrodes 135-140
over the \gls{vSMC}.

\clearpage

\subsection{ERPs across the vSMC}
\label{sec:erpsAcrossvSMC}

\begin{figure}
\begin{center}
\includegraphics[width=3.5in]{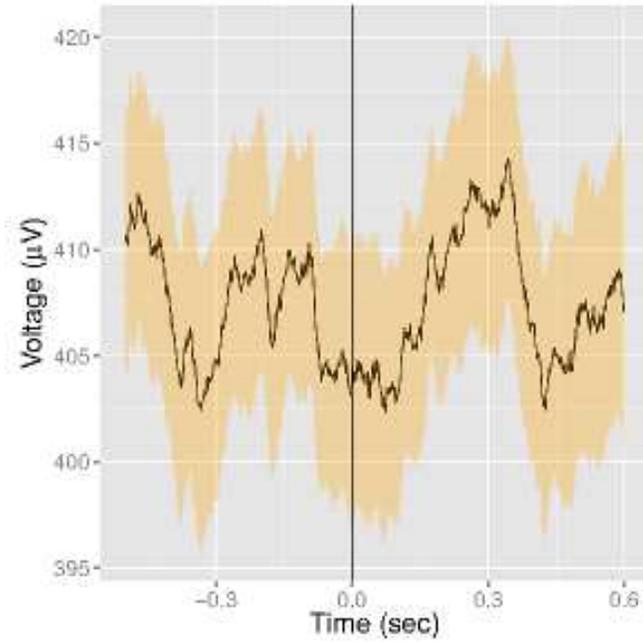}
\end{center}

\caption{ERP for electrode 129 computed from recordings in experimental
session EC2\_B105.}

\label{fig:erp129B105}
\end{figure}

\begin{figure}
\begin{center}
\includegraphics[width=3.5in]{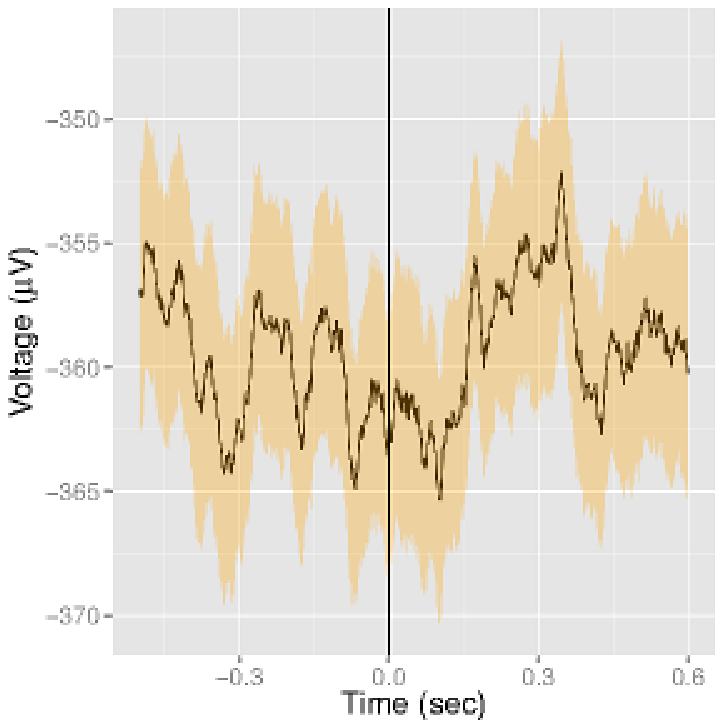}
\end{center}

\caption{ERP for electrode 130 computed from recordings in experimental
session EC2\_B105.}

\label{fig:erp130B105}
\end{figure}

\begin{figure}
\begin{center}
\includegraphics[width=3.5in]{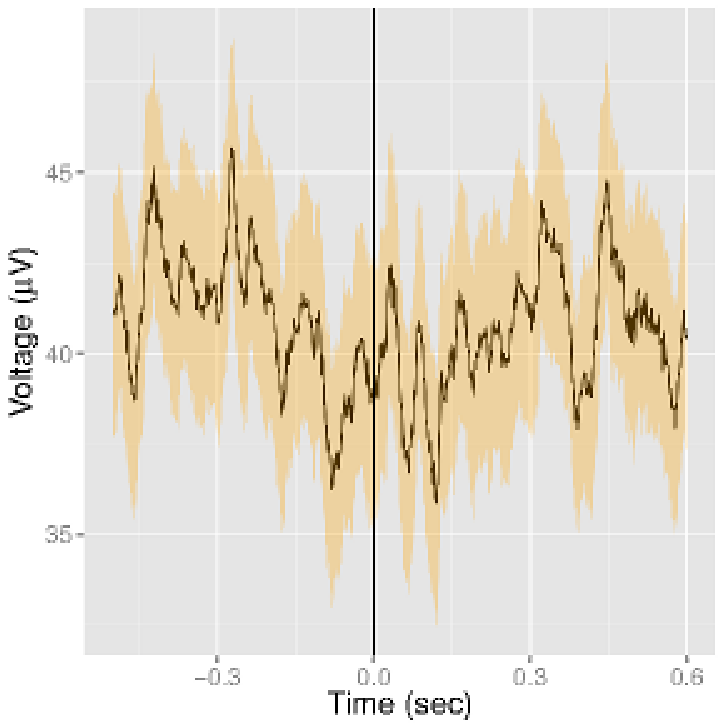}
\end{center}

\caption{ERP for electrode 131 computed from recordings in experimental
session EC2\_B105.}

\label{fig:erp131B105}
\end{figure}

\begin{figure}
\begin{center}
\includegraphics[width=3.5in]{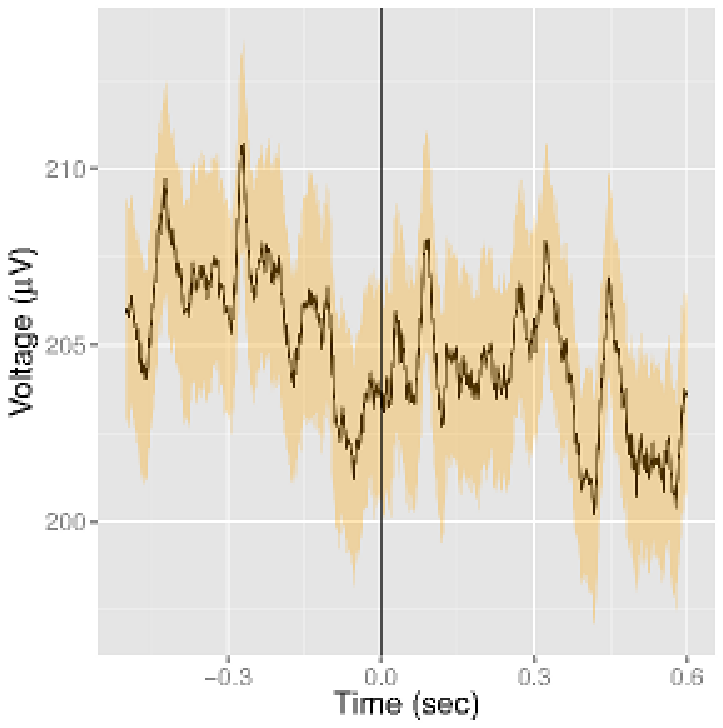}
\end{center}

\caption{ERP for electrode 132 computed from recordings in experimental
session EC2\_B105.}

\label{fig:erp132B105}
\end{figure}

\begin{figure}
\begin{center}
\includegraphics[width=3.5in]{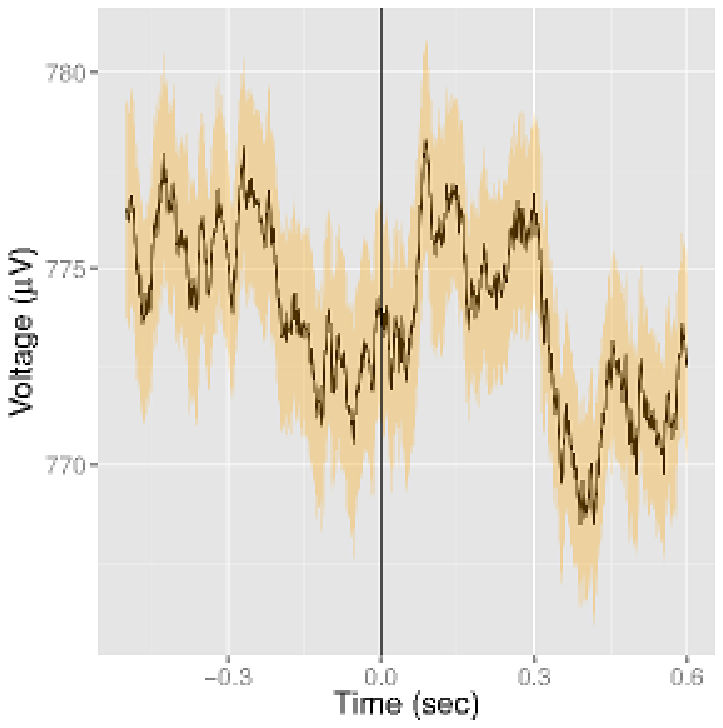}
\end{center}

\caption{ERP for electrode 133 computed from recordings in experimental
session EC2\_B105.}

\label{fig:erp133B105}
\end{figure}

\begin{figure}
\begin{center}
\includegraphics[width=3.5in]{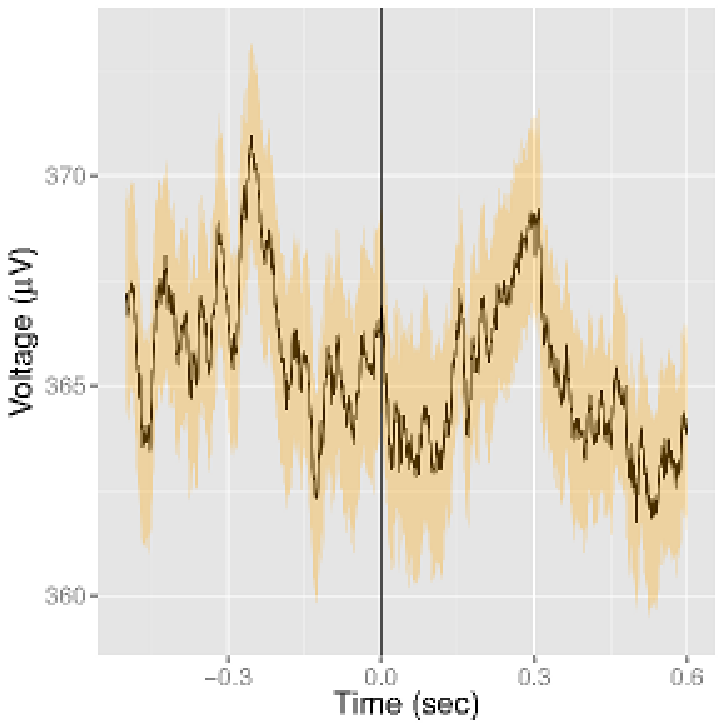}
\end{center}

\caption{ERP for electrode 134 computed from recordings in experimental
session EC2\_B105.}

\label{fig:erp134B105}
\end{figure}

\begin{figure}
\begin{center}
\includegraphics[width=3.5in]{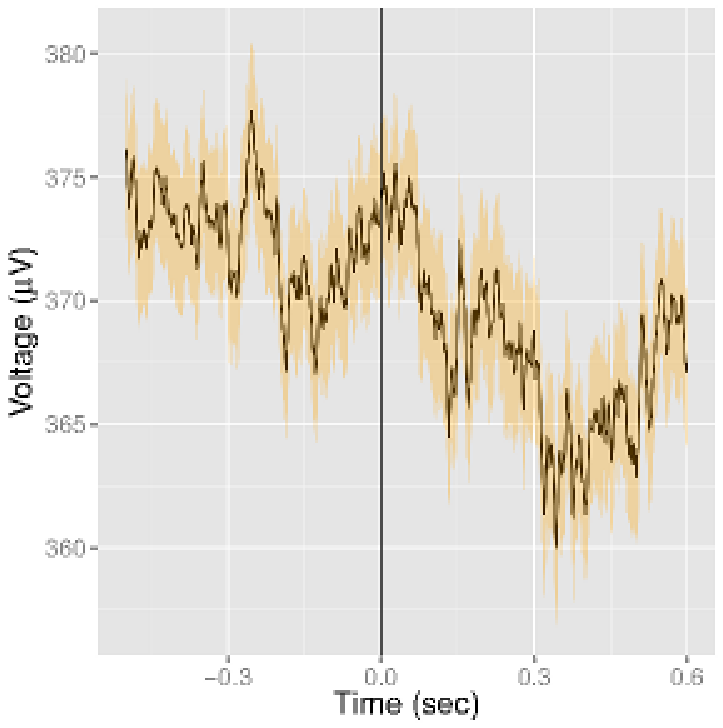}
\end{center}

\caption{ERP for electrode 135 computed from recordings in experimental
session EC2\_B105.}

\label{fig:erp135B105}
\end{figure}

\begin{figure}
\begin{center}
\includegraphics[width=3.5in]{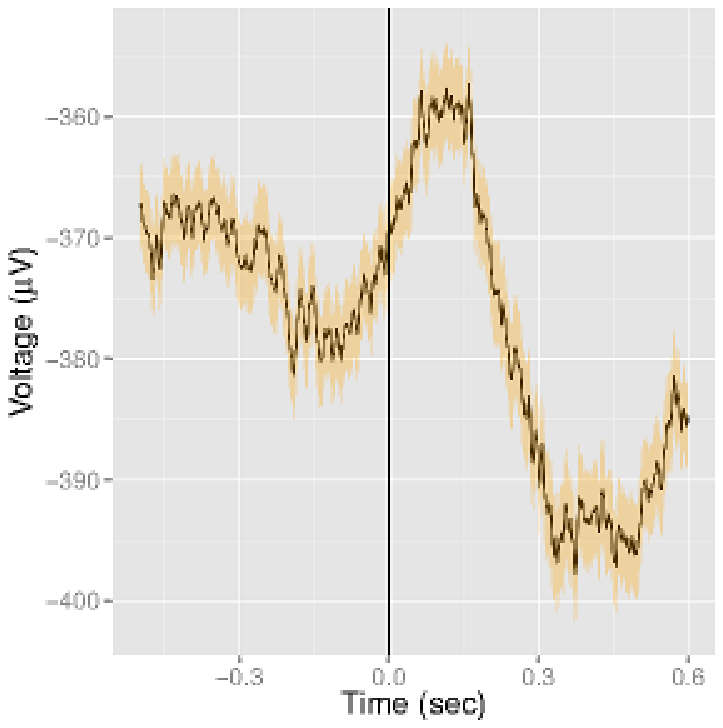}
\end{center}

\caption{ERP for electrode 136 computed from recordings in experimental
session EC2\_B105.}

\label{fig:erp136B105}
\end{figure}

\begin{figure}
\begin{center}
\includegraphics[width=3.5in]{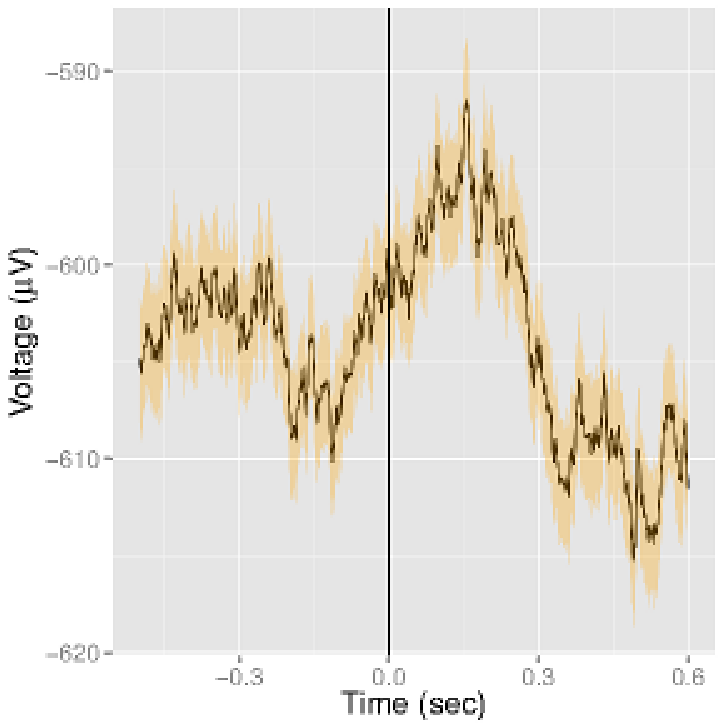}
\end{center}

\caption{ERP for electrode 137 computed from recordings in experimental
session EC2\_B105.}

\label{fig:erp137B105}
\end{figure}

\begin{figure}
\begin{center}
\includegraphics[width=3.5in]{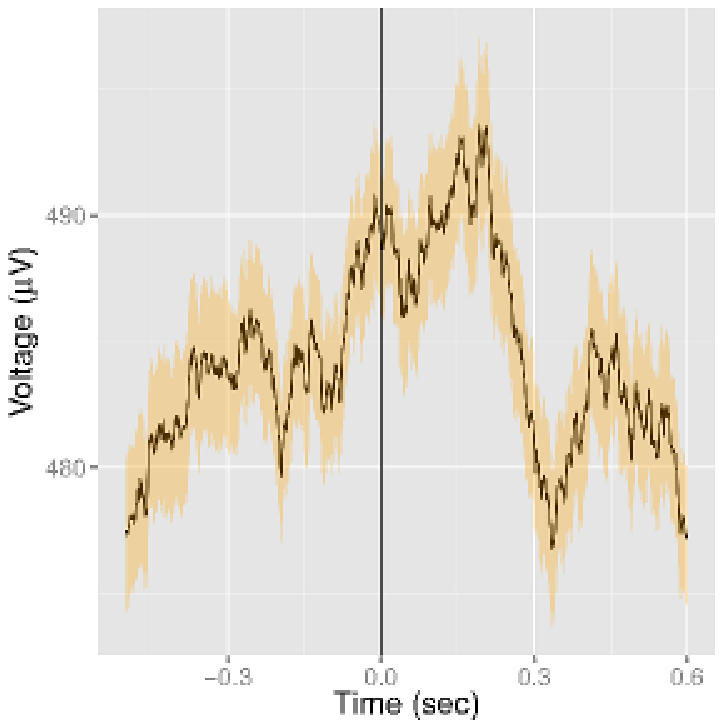}
\end{center}

\caption{ERP for electrode 138 computed from recordings in experimental
session EC2\_B105.}

\label{fig:erp138B105}
\end{figure}

\begin{figure}
\begin{center}
\includegraphics[width=3.5in]{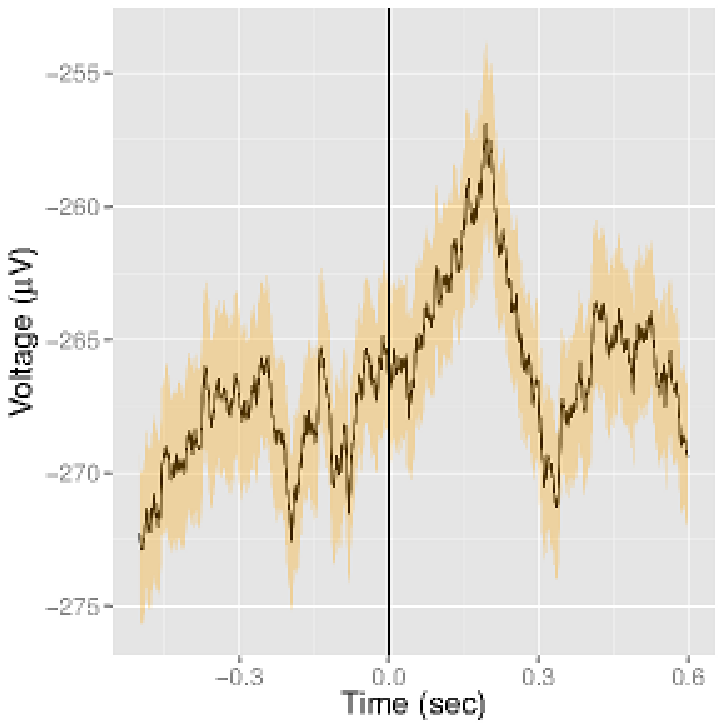}
\end{center}

\caption{ERP for electrode 139 computed from recordings in experimental
session EC2\_B105.}

\label{fig:erp139B105}
\end{figure}

\begin{figure}
\begin{center}
\includegraphics[width=3.5in]{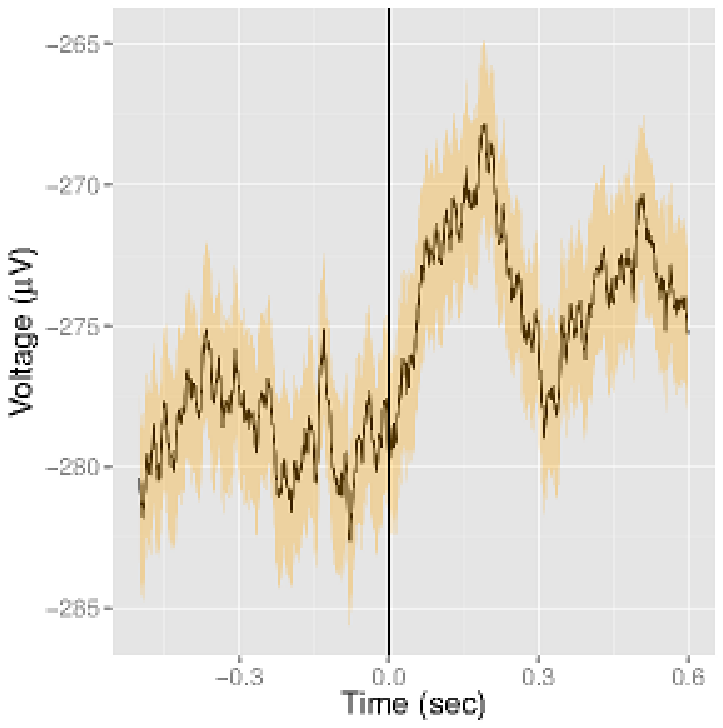}
\end{center}

\caption{ERP for electrode 140 computed from recordings in experimental
session EC2\_B105.}

\label{fig:erp140B105}
\end{figure}

\begin{figure}
\begin{center}
\includegraphics[width=3.5in]{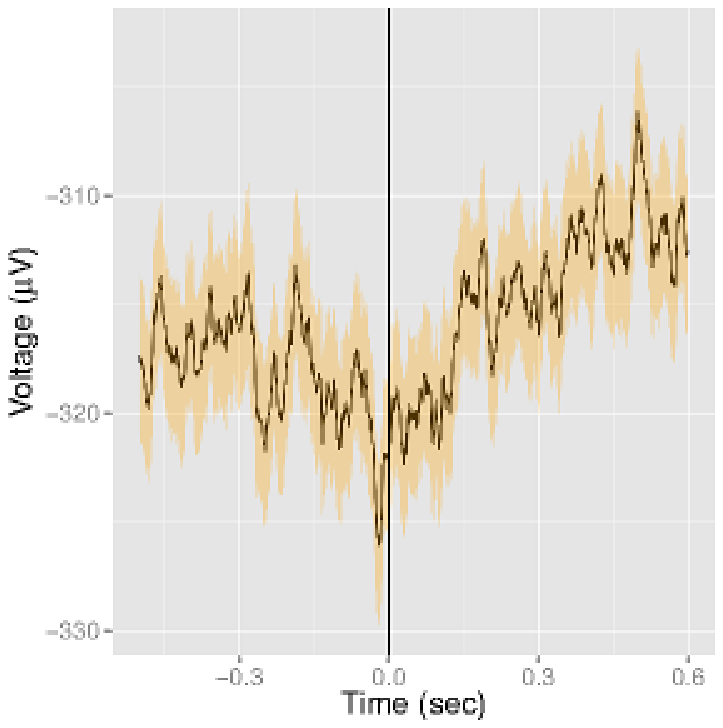}
\end{center}

\caption{ERP for electrode 141 computed from recordings in experimental
session EC2\_B105.}

\label{fig:erp141B105}
\end{figure}

Figures~\ref{fig:erp129B105}-\ref{fig:erp141B105} plots \glspl{ERP} computed
at electrodes along the ventro-dorsal axis of the recordings grid. Note that
\glspl{ERP} are larger over electrodes 136-140 in the \gls{vSMC}, in agreement
with Figure~\ref{fig:maxITCAcrossElectrodesB105}, and that the \gls{ERP} peak
is shifted to later times as we move from electrode 136 to electrode 140.

\subsection{Electrodes with strongest PAC}
\label{sec:strongestPAC}

\begin{figure}
\begin{center}
\includegraphics[width=6.0in]{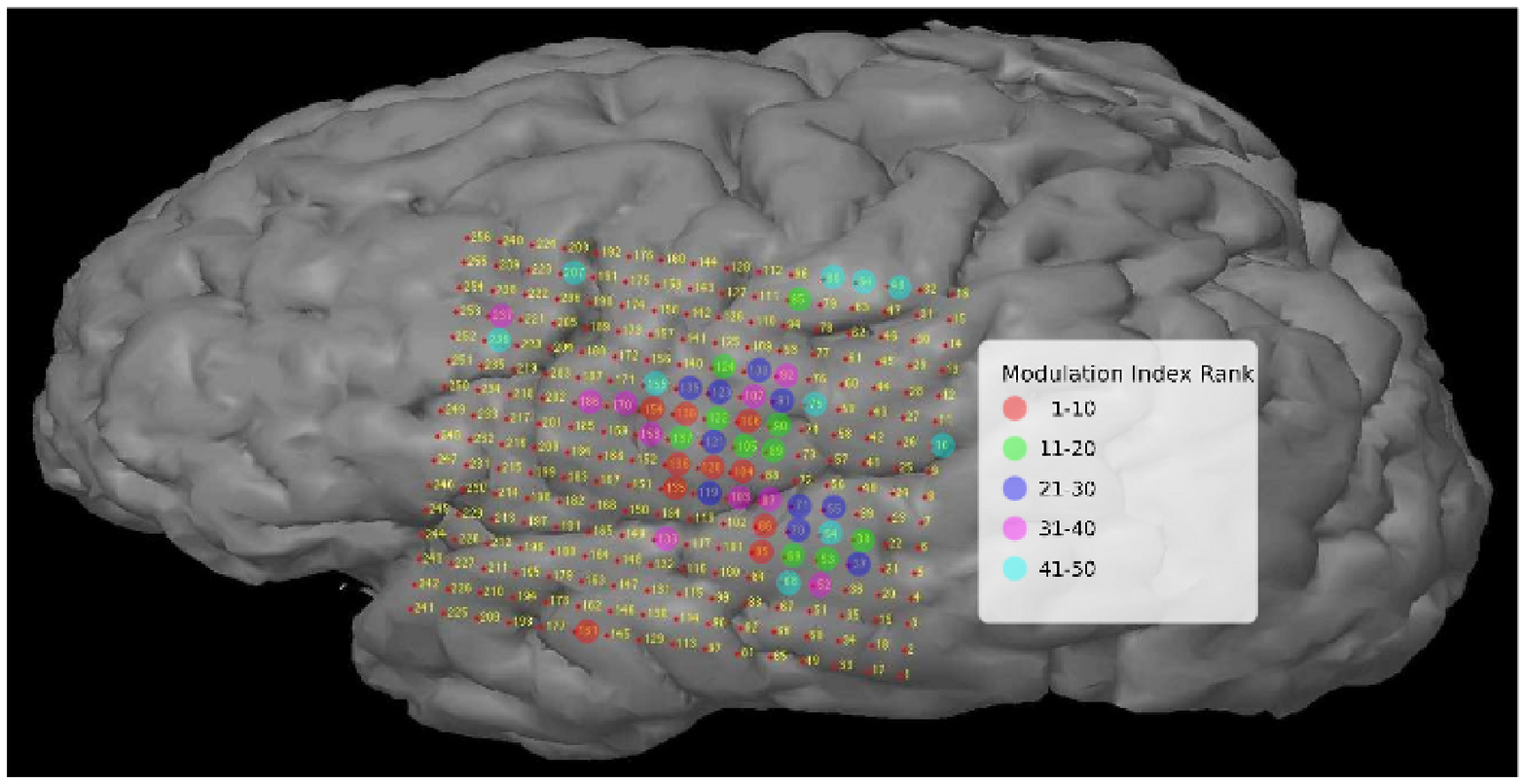}
\end{center}

\caption{Strongest \gls{PAC} occurs over the vSMC. The 50
electrodes with largest MI (between 0.5 seconds before and 0.6 seconds after
the CV transition, at the frequency of CV production, in the experimental
session EC2\_B105) are highlighted in color.}

\label{fig:largestMIsB105}
\end{figure}

Figure~\ref{fig:largestMIsB105} highlights the 50 electrodes with largest
\gls{MI}.

